\documentclass[a4paper,USenglish,cleveref]{lipics-v2021}
\usepackage{enumitem}

\hideLIPIcs
\nolinenumbers

\usepackage{siunitx}

 \category{} \supplement{Source code available at \url{https://github.com/koeppl/lzdr-comp}}

\title{LZD-style Compression Scheme with Truncation and Repetitions}
\author{Linus Götz}{TU Dortmund, Dortmund, Germany}{linus.goetz@tu-dortmund.de}{}{}{}
\author{Dominik K\"{o}ppl\footnote{Corresponding Author, dkppl@yamanashi.ac.jp}}{University of Yamanashi, Japan \and \url{https://dkppl.de/}}{dkppl@yamanashi.ac.jp}{https://orcid.org/0000-0002-8721-4444}{}\authorrunning{L.\ Götz and D.\ Köppl}
\Copyright{Linus Götz and Dominik Köppl}

\usepackage{amsmath,amsfonts,amssymb}

\usepackage[utf8]{inputenc}
\usepackage[T1]{fontenc}

\usepackage{enumerate}

\usepackage{url}

\usepackage[margin=0pt,font=small,labelfont=bf]{caption}

\usepackage[linesnumbered,ruled,vlined]{algorithm2e}
\makeatletter
\newcommand{\nosemic}{\renewcommand{\@endalgocfline}{\relax}}\newcommand{\dosemic}{\renewcommand{\@endalgocfline}{\algocf@endline}}\newcommand{\pushline}{\Indp}\newcommand{\popline}{\Indm\dosemic}\let\oldnl\nl \newcommand{\nonl}{\renewcommand{\nl}{\let\nl\oldnl}}\makeatother
\SetKw{Continue}{continue}
\SetKw{Break}{break}

\usepackage{float}
\usepackage{placeins}
\usepackage{subcaption}
\usepackage{adjustbox}
\usepackage{array}

\usepackage{forest}

\usepackage{tikz}
\usetikzlibrary{calc}
\usepackage{pgfplots}
\pgfplotsset{compat=1.18}
\usepgfplotslibrary{groupplots}

\newcommand*{\instancename}[1]{\ensuremath{\mathsf{#1}}} \newcommand*{\functionname}[1]{{{\renewcommand{\rmdefault}{ptm}\fontfamily{ppl}\selectfont\textrm{\textup{#1}}}}} 

\newcommand*{\ibeg}{\ensuremath{\textsf{b}}}
\newcommand*{\iend}{\ensuremath{\textsf{e}}}
\newcommand*{\succIndex}{\text{succ-index}}
\newcommand*{\fnSuccIndex}{\functionname{succ-index}}

\newcommand*{\LimitedLCE}{\functionname{LimitedLCE}}
\newcommand*{\fnLCE}{\functionname{LCE}}
\newcommand*{\factorIndex}{\functionname{index}}
\newcommand*{\child}{\functionname{child}}
\newcommand*{\fnLabel}{\functionname{label}}
\newcommand*{\fnPos}{\functionname{pos}}
\newcommand*{\fnLen}{\functionname{len}}

\newcommand*{\StdFlex}{\instancename{stdflex}}
\newcommand*{\AltFlex}{\instancename{altflex}}
\newcommand*{\MaxAlt}{\instancename{altmax}}
\newcommand*{\block}[1]{\noindent\textbf{#1.~}}

\begin{document}
\maketitle
\begin{abstract}
Lempel-Ziv-Double (LZD) is a variation of the LZ78 compression scheme 
that achieves better compression on repetitive datasets. 
Nevertheless, prior research has identified computational inefficiencies and a weakness in its compressibility for certain datasets.
In this paper, we introduce \emph{LZD+}, an enhancement of LZD, which enables expected linear-time online compression by allowing truncated references. 
To address the compressibility weakness exhibited by a lower bound example, 
we propose \emph{LZDR} (LZD-runlength compressed), a further enhancement on top of LZD+, 
which introduces a repetition-based factorization rule while maintaining linear expected time complexity for offline compression. 
Both time bounds can be de-randomized by a lookup data structure like a balanced search tree with a logarithmic dependency on the alphabet size. We prove that LZD+ and LZDR, like LZD, have an approximation ratio of $\Omega(n^{1/3})$ with respect to the smallest grammar, already on a single family of binary strings. Conversely, all three produce $O(n^{2/3}\delta^{1/3})$ factors, where $\delta$ is the substring complexity of the text.
Additionally, we present three flexible parsing variants of LZDR that yield fewer factors in practice; for one of them, we prove that it never yields more factors than greedy LZDR. 
Comprehensive benchmarking on standard corpora reveals that LZD+, LZDR, and its flexible variants outperform existing LZ-based methods in the number of factors while keeping competitive runtime efficiency.
However, we note that the difference in the number of factors becomes marginal for large datasets like those of the Pizza\&Chili corpus.
\end{abstract}

\section{Introduction}
Lempel--Ziv-78 (LZ78)~\cite{ziv78lz} is a data compression scheme that lies at the intersection between dictionary compression and grammar compression.
Several variations of LZ78 such as Lempel--Ziv--Welch (LZW)~\cite{welch84lzw} or LZMW~\cite{miller85variations} have been proposed to address specific shortcomings.
One of these variations is \emph{LZD (LZ-Double)}, which has been introduced by Goto et al.\ in 2015~\cite{goto15lzd}.
They showed that the compression ratio of LZD is better than LZ78 in most cases.
Given a text of length~$n$ whose letters are drawn from an alphabet of size~$\sigma$, 
Goto et al.\ provided an $O(n \log \sigma)$-time algorithm for computing LZD with suffix trees. 
They also provided an online algorithm using a dynamic trie to maintain the dictionary for selecting the next factor, which is a common approach to LZ78 and its variants.
However, Badkobeh et al.~\cite[Theorem~6]{badkobeh17two} have shown that the trie-based LZD factorization algorithm can exhibit a running time of $\Omega(n^{5/4})$ for input strings of length $n$.
As a remedy, Ohno et al.~\cite{ohno18lzabt} proposed a variant of LZD, which can be computed in $O(n \log n \log \sigma)$ time.
We are only aware of one linear-time algorithm for computing LZD, which however needs to preprocess the text to build heavy-weight data structures~\cite{koppl24computing}.

In this paper, we introduce another variant of LZD, called LZD+.
We show that LZD+ can be computed in linear expected time online by using a trie whose size is linear in the number of factors (\cref{thmLZDPlus}).
While each factor of an LZD factorization is the concatenation of two previous factors, LZD+ additionally supports the rules to select a prefix of the second factor, or to let a factor be the prefix of a previous factor.
These two additional rules make the dictionary of LZD+ prefix-closed, a feature present in LZ78 and LZW but absent in LZD and LZMW. This makes a trie-based online algorithm straightforward: every position on a path of the trie spells a prefix of a previous factor and hence a usable reference.

Furthermore, Badkobeh et al.\ have shown that there are strings of length $n$ for which the size of the grammars produced by the LZD factorization is larger than the size of the smallest grammar by a factor $\Omega(n^{\frac{1}{3}})$~\cite[Theorem~1]{badkobeh17two}.
We prove that this has not changed with LZD+ (\cref{thm:binary}), which is why we introduce another compression scheme called LZDR\@.
LZDR builds upon LZD+ and empirically avoids the bound of Badkobeh et al.\ while at the same time has an offline compression algorithm that runs in $O(n)$ expected time (\cref{thmLZDR}).
In comparison to LZD+, LZDR also allows a factor to be the repetition of a previous factor that can be truncated.
Moreover, we introduce three flexible parsing variants~\cite{matias01effect} of LZDR, which achieve fewer factors than (the greedy standard variant of) LZDR in practice.
For one of these variants, we prove that it never produces more factors than the greedy LZDR (\cref{prop:flex})\@.

Unfortunately, the improvement of LZDR on the strings of Badkobeh et al.\ is not universal.
Rotating the periodic part of these strings and replacing their separator letters by runs of~$\texttt{b}$ yields binary strings on which LZD, LZD+, and LZDR all produce $\Omega(n^{2/3})$ factors although a grammar of size $O(n^{1/3})$ exists (\cref{thm:binary}); hence, the approximation ratio of each of them is $\Omega(n^{1/3})$.
In the other direction, the counting argument by which Badkobeh et al.\ bounded the number of LZD factors extends to LZD+ and LZDR: each of the three greedy schemes produces $O(n^{2/3}\delta^{1/3})$ factors, where $\delta$ is the substring complexity of the text~\cite{kociumaka20towards}, a measure that is at most twice the size of a smallest grammar (\cref{thm:attractor}).
Moreover, LZDR produces at most one factor per maximal run of a single letter (\cref{thm:run}).

\begin{table}[t]
	\centering
	\caption{List of used symbols}
	\label{tab:symbols}
\begin{tabular}{ll}
  \textbf{Variable name} & \textbf{Meaning} \\
  $T$ & Input text \\
  $n$ & Length of $T$ \\
  $S$ & String \\
  $\ell$ & Length \\
  $\ell_x$ & Factor length \\
  $i,j$ & Text positions\\
  $x,y,w$ & Factor indices\\
  $z,\zeta$ & Number of factors in a factorization \\
  $z_{\mathrm{LZD}}, z_+, z_R$ & Numbers of factors of greedy LZD, LZD+, LZDR \\
  $g$ & Size of a smallest SLP generating $T$ \\
  $\delta$ & Substring complexity of $T$ \\
  $\operatorname{runs}(T)$ & Number of letter runs of $T$ \\
  $u, v$ & Radix trie nodes \\
  $e$ & Radix trie edge \\
  $\ibeg$ & Starting position of a factor \\
  $\iend$ & Ending position of a factor \\
  $F_x$ & Factor of a factorization \\
$R_x$ & Reference for flexible parsing \\
  $\mathcal{S}$ & Set of strings \\
\end{tabular}
\end{table}

\section{Preliminaries}\label{sec:preliminaries}
Our computational model is the word RAM model with words of $\Theta(\log n)$ bits; space bounds count words of working space and exclude the read-only text~$T$ and the output factorization.
A list of all defined symbols is given in \cref{tab:symbols}.

\block{Strings}
Let $\Sigma$ be a finite \emph{alphabet}. 
An element of $\Sigma$ is called a \emph{letter}, and an element of $\Sigma^*$ is called a \emph{string}.
The length of a string $S$ is denoted by $|S|$.
The empty string $\varepsilon$ is the string of length 0, i.e., $| \varepsilon | = 0$.
The concatenation of two strings $X$ and $Y$ is denoted by $X \cdot Y$ or $XY$.
For a string $S=XYZ$, $X$, $Y$ and $Z$ are called \emph{prefix}, \emph{substring} and \emph{suffix}, respectively.
The $i$th letter of a string $S$ is denoted by $S[i]$, where $1 \leq i \leq |S|$.
The substring of a string $S$ that starts at $i$ and ends at $j$ is denoted by $S[i..j]$, where $1 \leq i \leq j \leq |S|$.
Furthermore, we write the $i$th suffix of $S$ by $S[i..] = S[i..|S|]$.
The $k$-fold \emph{repetition} of a string $S$ is denoted by $S^k$ and is defined inductively as $S^0 = \varepsilon$ and $S^{k+1} = S^k \cdot S$.
The infinitely repeated string $S^\infty$ is defined as $S^\infty = \lim_{k \rightarrow \infty} S^k$.

\block{Trie}
A \emph{radix trie} is a trie whose unary paths are compacted.
In detail, a radix trie represents a set $\mathcal{S}$ of strings in form of a rooted tree that satisfies the following properties:
	\begin{enumerate}[wide, labelwidth=!, labelindent=0pt]
    \item Each edge~$e$ has an associated non-empty string that is a substring of a string in $\mathcal{S}$.
			We call this associated string of $e$ the \emph{edge label} and denote it by $e.\fnLabel{}$.
			Further, we write $|e| = |e.\fnLabel{}|$ as a shorthand.
    \item For any two distinct outgoing edges of the same node, the edge labels of these edges have to start with distinct letters.
		\item 
		    The \emph{string label} of a node $u$ is the concatenation of the edge labels of all edges visited when traversing from the root to $u$.
		    For each string $S \in \mathcal{S}$, there is a node $v$ whose string label is $S$.
\end{enumerate}
We define the \emph{start node} (resp.\ \emph{end node}) of an edge as the node that is connected to the edge and is the closest to (resp.\ the furthest from) the root node.
We call a node~$u$ a \emph{factor node} if $u$'s string label is an element of $\mathcal{S}$.
Otherwise, we call $u$ a \emph{split node}.

For the remaining part, we fix a string $T$ of length $n$ whose letters are drawn from an integer alphabet $\Sigma$ with $\sigma := |\Sigma| \ge 2$.
We call $T$ the \emph{text}.
For technical reasons, define $T[n+1..]$ to be the empty string.
We define two queries on $T$:
First, the \emph{longest common extension} (LCE) query $T.\fnLCE(i, j)$ for $T$ returns the length of the longest common prefix of $T[i..]$ and $T[j..]$.
For example, if $T = \texttt{babababb}$, the LCE query $T.\fnLCE(2, 4)$ returns $4$ because the longest common prefix is $\texttt{abab}$.
Second, we introduce a function $T.\LimitedLCE(i,j,\ell)$ for $T$ that returns the length of the longest common prefix of $T[i..i+\ell-1]$ and $T[j..j+\ell-1]$. 
We implement this function by naive letter-wise comparisons, which take $O(\ell)$ time.

\block{Factorizations}
A \emph{factorization} of $T$ is a list of substrings $F_1, F_2, \ldots, F_z$ of $T$ such that $T = F_1 F_2 \cdots F_z$.
The \emph{factor count} of the factorization is the number $z$ of substrings.
Each substring $F_x$ is called a \emph{factor}. 
With \emph{factor index}, we refer to the number $x$ of a factor $F_x$.
We let $\ibeg(F_x)$ and $\iend(F_x)$ denote, respectively, the starting and ending position of $F_x$ in $T$ such that
$F_x = T[\ibeg(F_x)..\iend(F_x)]$ and $\iend(F_{x-1}) + 1 = \ibeg(F_x)$ for every $x \in [2..z]$.
For convenience, we stipulate that $F_0 = \varepsilon$ always denotes the string of length 0 with $\ibeg(F_0) = \iend(F_0) = 0$. An algorithm computing a factorization is \emph{online} if it processes $T$ from left to right without preprocessing~$T$, and it has read only $T[1..\iend(F_x)+1]$ when it outputs~$F_x$.

The \emph{LZD factorization}~\cite{goto15lzd} of $T$ is the factorization $T = F_1 F_2 \cdots F_z$ such that, for every $x \in [1..z]$, 
$F_x = F_{x_1} F_{x_2}$ where $F_{x_1}$ is the longest prefix of $T[\ibeg(F_x)..n]$ with $F_{x_1} \in \{ F_1, \ldots, F_{x-1} \} \cup \Sigma$, and $F_{x_2}$ is the longest prefix of $T[\ibeg(F_x)+|F_{x_1}|..n]$ with $F_{x_2} \in \{ F_0, F_1, \ldots, F_{x-1} \} \cup \Sigma$.

The \emph{dictionary} is a dynamic set~$\mathcal{S}$ of factors that can be referenced to build the next factor.
For example, the dictionary for LZD at the point of building the factor $F_x$ is $\mathcal{S} = \{ F_0, F_1, \ldots, F_{x-1} \}$.
If a factor $F_x$ builds upon a previous factor $F_y$ with $y \in [1..x-1]$ and references it, we call $F_y$ the \emph{reference}.
We call a dictionary \emph{prefix-closed} if all prefixes of each element (which is a string) of the dictionary can be used to form $F_x$.
Unlike the variants we introduce next, the dictionary of LZD is not prefix-closed~\cite[Figure~2]{koppl24computing}.

\block{Grammars}
A \emph{straight-line program} (SLP) is a context-free grammar in which every nonterminal has exactly one rule, of the form $X \to c$ for a letter~$c$ or $X \to YZ$ for two nonterminals $Y$ and~$Z$ defined before~$X$; the SLP generates the string derived from its last nonterminal.
We denote by~$g$ the minimum number of rules of an SLP generating~$T$; up to a constant factor, $g$ is also the size of a smallest context-free grammar generating only~$T$, which we call the \emph{smallest grammar}.
Intuitively, an SLP stores each repeated building block only once.
Since a rule at most doubles the length of the longest string derived so far, $n \le 2^{g-1}$, i.e., $g = \Omega(\log n)$.
An LZD factorization with $z$ factors induces an SLP with $O(z)$ rules, which we call the \emph{grammar produced by LZD}: one rule $X_x \to X_{x_1}X_{x_2}$ per factor, one rule per letter, and $z-1$ rules concatenating the factors.
For a factorization scheme producing $z(T)$ factors on~$T$, we say that its \emph{approximation ratio} (with respect to the smallest grammar) is 
\begin{itemize}
  \item $\Omega(f(n))$ if there are infinitely many texts~$T$ with $z(T)/g(T) = \Omega(f(|T|))$, and 
  \item $O(f(n))$ if $z(T)/g(T) = O(f(|T|))$ holds for all texts~$T$.
\end{itemize}

\begin{algorithm}[t]
  \scriptsize
  \DontPrintSemicolon
  \caption{Finding the longest reference or return a single letter}
  \label{alg:longestref}
  \KwData{Entire input string $T$ of length $n$, starting position for current factor $k \in \mathbb{N}^+$, $previousFactors$ as radix trie}
  \SetKwProg{Def}{def}{:}{}
  \SetKwFunction{FLongestRef}{GetLongestFactorRef}
  \Def{\FLongestRef{$T$, $k$, previousFactors}}{
    \tcp{Initialize the longest factor with the empty factor $F_0$}
    $y \gets \{ factor \gets 0, \ length \gets 0 \}$ \;
    \tcp{Initialize the current node and current index}
    $u \gets \text{root node of $previousFactors$}$ \;
    $i \gets 0$ \; \label{alg:longestref:initindex}
    \While{$k + i \leq n$}{
      \tcp{Get end node of edge that starts with the current letter}
      $v \gets u.\child(T[k + i])$ \;
      \If{$v = \bot$}{
        \tcp{Edge does not exist}
        \tcp{We were unable to read the letter, therefore break loop}
        \Break \;
      }
      \tcp{Get edge from $u$ to $v$}
      $e \gets (u, v)$ \;
\tcp{Find length of longest common prefix of $T[k+i..]$ and edge label of $e$}
      $commonLength \gets T.\LimitedLCE(k + i, e.\fnPos, e.\fnLen)$ \;
      \label{alg:longestref:edgelce}

      \eIf{$commonLength \geq e.\fnLen$}{
        \tcp{edge label successfully read, increase $i$}
        $i \gets i + e.\fnLen$ \;
        \tcp{Go to next node}
        $u \gets v$ \;
        \If{$\text{$u$ is not split node}$}{ \label{alg:longestref:updatefactor}
          \tcp{Update factor}
          $y \gets \{ factor \gets \factorIndex(u), \ length \gets i \}$ \;
        }
      }{
        \tcp{Mismatch during edge label of $e$, could not reach next node}
        \Break \;
      }
    }

    \tcp{Use single letter if no matching factor found, or when factor has length $1$}
    \If{$k \leq n$ $\boldsymbol{and}$ $y.length \leq 1$}{
      $y \gets \{ factor \gets T[k], \ length \gets 1 \}$ \;
    }

    \Return $y$ \;
  }
\end{algorithm}

\begin{algorithm}[t]
  \scriptsize
  \DontPrintSemicolon
  \caption{Finding the next longest LZD factor}
  \label{alg:nextlzdfactor}
  \KwData{Entire input string $T$ of length $n$, starting position for current factor $k \in \mathbb{N}^+$, $previousFactors$ as radix trie}
  \KwResult{the next longest LZD factor}
  \SetKwProg{Def}{def}{:}{}
  \SetKwFunction{FLongestLZDFactor}{NextLongestLzdFactor}
  \SetKwFunction{FLongestRef}{GetLongestFactorRef}
  \SetKwFunction{FLongestTruncation}{GetLongestFactorTruncation}
  \SetKwFunction{FLongestRepetition}{GetLongestFactorRepetition}
  \Def{\FLongestLZDFactor{$T$, $k$, previousFactors}}{
    \tcp{Combination factor (\Cref{alg:longestref})}
    $cf1 \gets \FLongestRef{T, k, previousFactors}$ \;
    $cf2 \gets \FLongestRef{T, $k + cf1.length$, previousFactors}$ \;
    \nosemic $cf \gets \{$ \;
      \pushline\nosemic $firstFactor \gets cf1.factor$, \;
      \nosemic $secondFactor \gets cf2.factor$, \;
      \nosemic $length \gets cf1.length + cf2.length$ \;
    \popline\nosemic $\}$ \;
    \Return $cf$ \;
  }
\end{algorithm}

\section{LZD+ compression scheme}\label{sec:lzd}
We propose a greedy compression scheme, called LZD+, that is based on LZD with two modifications.
First, we modify the combination rule $F_x = F_{x_1} F_{x_2}$ to allow selecting a prefix of the second reference~$F_{x_2}$.
Second, we introduce another factor production rule besides combinations: selecting a prefix of a previous factor, which we call \emph{truncation}.
These modifications have the advantage that, unlike LZD, there is a straightforward compression algorithm for LZD+, which uses a radix trie and runs in linear expected time, 
which does not seem easy for LZD\@.

A trie-based algorithm for LZD has the problem that while we are searching for a reference (\Cref{alg:nextlzdfactor,alg:longestref}) and are descending the trie to find the deepest factor node whose string label is a prefix of the remaining input, 
there is no guarantee that we will reach any factor node.
In detail, \cref{alg:nextlzdfactor} summarizes the computation of the next LZD factor using \cref{alg:longestref}.
We first find the longest complete reference starting at the current position, and then the longest complete reference immediately following it. 
Their concatenation forms the next LZD factor. 

\block{Definition of LZD+}
The \emph{LZD+ factorization} of $T$ is the factorization $T = F_1 F_2 \cdots F_z$, 
such that, for every $x \in [1..z]$, the two possible production rules for $F_x$ are:
\begin{enumerate}
    \item Combination: $p_1 = \left( F_{x_1} F_{x_2} \right) [1..\ell_x]$ where $p_1$ is a prefix of $T[\ibeg(F_x)..n]$ and $F_{x_1}$ is the longest prefix of $T[\ibeg(F_x)..n]$ with $F_{x_1} \in \{ F_1, \ldots, F_{x-1} \} \cup \Sigma$, $F_{x_2} \in \{ F_0, F_1, \ldots, F_{x-1} \} \cup \Sigma$, and $1 \leq \ell_x \leq |F_{x_1}| + |F_{x_2}|$.
    \item Truncation: $p_2 = F_y[1..\ell_x]$ where $p_2$ is a prefix of $T[\ibeg(F_x)..n]$ with $F_y \in \{ F_1, \ldots, F_{x-1} \}$, and $1 \leq \ell_x \leq |F_y|$.
\end{enumerate}
The factor $F_x$ is then chosen among all possible $p_1$'s and $p_2$'s as the one that gives the maximum length. 
If $F_{x_1}$ or $F_{x_2}$ have length~1, we prefer to use a single letter instead of a previous factor.

Intuitively, a combination glues the longest previous factor (or letter)~$F_{x_1}$ that is a prefix of the remaining text to a prefix of a second previous factor, as in LZD but with the second reference possibly truncated, while a truncation copies a prefix of a single previous factor~$F_y$.
We call a previous factor or letter a \emph{complete reference} at a position if it is a prefix of the text starting there, in contrast to a truncated reference, which is only required to have a prefix matching the text.
The truncation rule is not redundant: if, for example, \texttt{abcabc} is the only previous factor starting with~\texttt{a} and no previous factor starts with~\texttt{b}, then at the text \texttt{abcabx} the truncation yields \texttt{abcab}, whereas the longest combination is \texttt{ab}.
Since the combination rule allows every length $\ell_x \in [1..|F_{x_1}|+|F_{x_2}|]$ and the truncation rule every length $\ell_x \in [1..|F_y|]$, the lengths of the admissible candidates at a fixed position form an interval starting at~$1$; we use this in \cref{sec:flexible}.

\begin{table}[H]
  \scriptsize
  \centering
  \begin{tabular}{|c|c|c|c|c|c|c|c|c|c|c|}
    \hline
    $\boldsymbol{F_1}$ & $\boldsymbol{F_2}$ & $\boldsymbol{F_3}$ & $\boldsymbol{F_4}$ & $\boldsymbol{F_5}$ & $\boldsymbol{F_6}$ & $\boldsymbol{F_7}$ & $\boldsymbol{F_8}$ & $\boldsymbol{F_9}$ & $\boldsymbol{F_{10}}$ & $\boldsymbol{F_{11}}$ \\\hline
    aa & bb & aabb & baabb & bbba & ba & baab & cc & ccba & bab & c \\\hline
    $(\texttt{a}, \texttt{a})$ & $(\texttt{b}, \texttt{b})$ & $(F_1, F_2)$ & $(\texttt{b}, F_3)$ & $(F_2, F_4)[1..4]$ & $(\texttt{b}, \texttt{a})$ & $F_4 [1..4]$ & $(\texttt{c}, \texttt{c})$ & $(F_8, F_6)$ & $(F_6, \texttt{b})$ & $(\texttt{c}, F_0)$ \\\hline
  \end{tabular}
  \caption{LZD+ factorization of the string \texttt{aabbaabbbaabbbbbababaabccccbababc}.
In practice, we can represent the factors as
$F_1 = (\texttt{a}, \texttt{a})$,
$F_2 = (\texttt{b}, \texttt{b})$,
$F_3 = (F_1, F_2)$,
$F_4 = (\texttt{b}, F_3)$,
$F_5 = (F_2, F_4)[1..4]$,
$F_6 = (\texttt{b}, \texttt{a})$,
$F_7 = F_4 [1..4]$,
$F_8 = (\texttt{c}, \texttt{c})$,
$F_9 = (F_8, F_6)$,
$F_{10} = (F_6, \texttt{b})$,
and $F_{11} = (\texttt{c}, F_0)$.
  }
	\label{tab:lzdplusEx}
\end{table}

\begin{example} \label[example]{ex:lzdplusEx}
The LZD+ factorization of the string \texttt{aabbaabbbaabbbbbababaabccccbababc} is 
$F_1 = \texttt{aa}$ (Combination of $\texttt{a}$ and $\texttt{a}$),
$F_2 = \texttt{bb}$ (Combination of $\texttt{b}$ and $\texttt{b}$),
$F_3 = \texttt{aabb}$ (Combination of $F_1$ and $F_2$),
$F_4 = \texttt{baabb}$ (Combination of $\texttt{b}$ and $F_3$),
$F_5 = \texttt{bbba}$ (Combination of $F_2$ and $F_4$ with truncation applied),
$F_6 = \texttt{ba}$ (Combination of $\texttt{b}$, and $\texttt{a}$ or truncating $F_1$),
$F_7 = \texttt{baab}$ (Truncation of $F_4$),
$F_8 = \texttt{cc}$ (Combination of $\texttt{c}$ and $\texttt{c}$),
$F_9 = \texttt{ccba}$ (Combination of $F_8$ and $F_6$),
$F_{10} = \texttt{bab}$ (Combination of $F_6$ and $\texttt{b}$),
and $F_{11} = \texttt{c}$ (Combination of $\texttt{c}$ and $F_0$).
We visualized the factorization in \cref{tab:lzdplusEx}. The factors $F_5$ and~$F_7$ illustrate the two uses of truncation: $F_5$ truncates the second reference of a combination, whereas $F_7$ is a standalone truncation of~$F_4$.
\end{example}

\block{Augmented radix trie}
We use the radix trie described in \cref{sec:preliminaries} to dynamically maintain all computed factors.
To obtain a linear-time algorithm for LZD+, we augment the trie with some data, which we describe in the following.
First, we want to compute a function $\factorIndex(u)$ that returns the index represented by a node $u$ in $O(1)$ time.
For that, we let a factor node representing $F_x$ store the factor index~$x$.
Further, we let the root node and all split nodes store the factor index~$0$.

Second, we want to compute a function $u.\child(c)$ that, for a node $u$, returns in $O(1)$ expected time the end node of the outgoing edge from $u$ whose edge label starts with $c$.
However, the function $\child$ returns $\bot$ if no such edge exists.
For that, we implement $\child$ by a hash table representing an associative array storing pairs of the form $(u,c)$ as keys.

Finally, we need a mechanism to support truncations efficiently.
Suppose we need to find a reference that has a prefix matching the string label of $u$ concatenated with $c$.
Further, assume that the node $v = u.\child(c)$ exists, and there is a mismatch between the remaining text and the string label of $v$.
Our goal is therefore to find any factor index stored in the subtree rooted at $v$.
While we can access $v$ via $\child$, $v$ might actually be a split node instead of a factor node, and thus is not helpful.
Indeed, a worst case is to traverse downwards to a leaf while visiting many split nodes.
As a remedy, we let each node~$u$ save an additional number, which we call the \textit{\succIndex{}}, which is the factor index of any factor node in the subtree rooted in $u$.
For a factor node~$u$, this number is $\factorIndex(u)$.
For the split nodes, we set their \succIndex{} during their creation.
We create a split node only when inserting a factor~$F_x$ into the radix trie.
At that time, the factor node representing $F_x$ is a child of the created split node~$u$, and thus it suffices to set the \succIndex{} of~$u$ to $x$.

Moreover, to get the radix trie into $O(z)$ space for storing $z$ factors, 
we do not store an edge label $e.\fnLabel$ in plain form but represent it by a pointer $e.\fnPos$
and a length $e.\fnLen$ such that $e.\fnLabel = T[e.\fnPos..e.\fnPos+e.\fnLen-1]$.
\Cref{fig:radixtrieLZDP} depicts an example of our radix trie.

\begin{figure}
  \hspace{0.05\textwidth}
  \begin{subfigure}[t]{0.18\textwidth}
    \centering
    \begin{forest}
      for tree={circle,draw, l sep=20pt, minimum size=32pt, inner sep=0pt}
      [0(0)
        [1(1),double,edge label={node[midway,left] {aa}} ]
      ]
    \end{forest}
    \caption{After insertion of $F_1 = \texttt{aa}$}
    \label{fig:radixtrieLZDPA}
  \end{subfigure}
  \hspace{0.05\textwidth}
  \begin{subfigure}[t]{0.18\textwidth}
      \centering
      \begin{forest}
        for tree={circle,draw, l sep=20pt, minimum size=32pt, inner sep=0pt}
        [0(0)
          [1(1),double,edge label={node[midway,left] {aa}} ]
          [2(2),double,edge label={node[midway,right] {bb}} ]
        ]
      \end{forest}
      \caption{After insertion of $F_2 = \texttt{bb}$}
      \label{fig:radixtrieLZDPB}
  \end{subfigure}
  \hspace{0.05\textwidth}
  \begin{subfigure}[t]{0.40\textwidth}
      \centering
      \begin{forest}
        for tree={circle,draw, l sep=20pt, minimum size=32pt, inner sep=0pt}
        [0(0)
          [1(1),double,edge label={node[midway,left] {aa}}, tier=d1
            [3(3),double,edge label={node[midway,left] {bb}}, tier=d2 ]
          ]
          [0(2),edge label={node[midway,right] {b}}, tier=d1
            [6(6),double,edge label={node[midway,left] {a}}, tier=d2
              [7(7),double,edge label={node[midway,left] {ab}}, tier=d3
                [4(4),double,edge label={node[midway,left] {b}}, tier=d4 ]
              ]
              [10(10),font=\scriptsize,double,edge label={node[midway,right] {b}}, tier=d3 ]
            ]
            [2(2),double,edge label={node[midway,right] {b}}, tier=d2
              [5(5),double,edge label={node[midway,right] {ba}}, tier=d3 ]
            ]
          ]
          [11(11),font=\scriptsize,double,edge label={node[midway,right] {c}}, tier=d1
            [8(8),double,edge label={node[midway,right] {c}}, tier=d2
              [9(9),double,edge label={node[midway,right] {ba}}, tier=d3 ]
            ]
          ]
        ]
      \end{forest}
      \caption{After all factors from $F_1$ to $F_{11}$ have been inserted}
      \label{fig:radixtrieLZDPC}
  \end{subfigure}
  \caption{Radix trie for LZD+ factorization of \texttt{aabbaabbbaabbbbbababaabccccbababc}. Double circled nodes represent factor nodes, while single circled nodes represent split nodes. The first number represents the index of the factor node, and the number in parentheses is the \succIndex{}.}
  \label{fig:radixtrieLZDP}
\end{figure}

\begin{algorithm}[t]
  \scriptsize
  \DontPrintSemicolon
  \caption{Finding the longest truncation of a previous factor.
We define a function $\protect\fnSuccIndex(u)$, that returns the \succIndex{} for a node $u$ in $O(1)$ time.
  }
  \label{alg:longesttruncation}
  \KwData{Entire input string $T$ of length $n$, starting position for current factor $k \in \mathbb{N}^+$, $previousFactors$ as radix trie}
  \SetKwProg{Def}{def}{:}{}
  \SetKwFunction{FLongestTruncation}{GetLongestFactorTruncation}
  \Def{\FLongestTruncation{$T$, $k$, previousFactors}}{
    \tcp{Initialize the longest factor with the empty factor $F_0$}
    $y \gets \{ factor \gets 0, \ length \gets 0 \}$ \;
    \tcp{Initialize the current node and current index}
    $u \gets \text{root node of $previousFactors$}$ \;
    $i \gets 0$ \;
    \While{$k + i \leq n$}{ \label{alg:longesttruncation:whiledef}
      \tcp{Get end node of edge that starts with the current letter}
      $v \gets u.\child(T[k + i])$ \;
      \If{$v = \bot$}{
        \tcp{Edge does not exist}
        \tcp{We were unable to read the letter, therefore break loop}
        \Break \;
      }
      \tcp{Get edge from $u$ to $v$}
      $e \gets (u, v)$ \;
      \tcp{Find length of longest common prefix of $T[k+i..]$ and edge label of $e$}
      $commonLength \gets T.\LimitedLCE(k + i, e.\fnPos, e.\fnLen)$ \;
      \eIf{$commonLength \geq e.\fnLen$}{
        \tcp{edge label successfully read, increase $i$}
        $i \gets i + e.\fnLen$ \;
        \tcp{Go to next node}
        $u \gets v$ \;
        \tcp{Update factor}
        $y \gets \{ factor \gets \fnSuccIndex(u), \ length \gets i \}$ \; \label{alg:longesttruncation:alwaysupdate}
      }{
        \tcp{Mismatch during edge label of $e$, could not reach next node}
        $i \gets i + commonLength$ \; \label{alg:longesttruncation:extraupdate}
        \tcp{Update factor}
        $y \gets \{ factor \gets \fnSuccIndex(v), \ length \gets i \}$ \;
        \Break \;
      }
    }
    \Return $y$ \;
  }
\end{algorithm}

\block{Factorization algorithm}
In the following, we describe the process for constructing an LZD+ factorization of $T$.
We build the factorization by incrementally searching for the next factor $F_x$ based on the remaining input with the dictionary storing all previous factors $F_1, \ldots, F_{x-1}$.
To this end, we use two helper functions for computing $F_x$, and later show how we will use them.
\begin{enumerate}[label=(\alph*)]
    \item Find the longest reference or return a single letter (\Cref{alg:longestref}). \label{it:longRef}
    \item Find the longest truncation (of a previous factor) (\Cref{alg:longesttruncation}). \label{it:Truncation}
\end{enumerate}

In a nutshell, computing~$F_x$ takes three traversals of the radix trie along the remaining text: one for the longest standalone truncation, one for the longest complete reference~$F_{x_1}$, and one, starting directly after~$F_{x_1}$, for the longest prefix of a second reference.
A traversal for a complete reference remembers the last factor node it reaches.
A traversal for a truncation may also end at a split node~$u$ or inside an edge~$(u,v)$; then every factor in the subtree of~$u$, respectively of~$v$, has the matched string as a prefix, and the \succIndex{} of this node names one such factor.
This is the only additional information needed to make all prefixes of previous factors usable.

For both helper functions, we first initialize variables that aid us with the traversal of the trie.
\begin{enumerate}
  \item A pointer $u$ to the current node that is initialized with the root node of the trie.
  \item An integer $i$ equal to the number of letters that have been read from the remaining input, initialized to 0 (e.g., at Line~\ref{alg:longestref:initindex} in \cref{alg:longestref}).
\end{enumerate}
We also maintain a variable $y$ that stores the index of the factor with the (currently so far found) longest prefix matching the remaining text.
We call $y$ the (index of the) \emph{reference candidate}, and initialize $y$ to 0, the index of the empty factor~$F_0$.
(Returning $F_0$ is the correct choice if the remaining input is empty.)

\begin{algorithm}
  \scriptsize
  \DontPrintSemicolon
  \caption{Finding the next longest LZD+ factor in $O(\ell)$ time}
  \label{alg:nextlzdpfactor}
  \KwData{Entire input string $T$ of length $n$, starting position for current factor $k \in \mathbb{N}^+$, $previousFactors$ as radix trie}
  \KwResult{the next longest LZD+ factor of length $\ell$}
  \SetKwProg{Def}{def}{:}{}
  \SetKwFunction{FLongestLZDPFactor}{NextLongestLzdpFactor}
  \SetKwFunction{FLongestRef}{GetLongestFactorRef}
  \SetKwFunction{FLongestTruncation}{GetLongestFactorTruncation}
  \SetKwFunction{FLongestRepetition}{GetLongestFactorRepetition}
  \Def{\FLongestLZDPFactor{$T$, $k$, previousFactors}}{
    \tcp{Combination factor (\Cref{alg:longestref,alg:longesttruncation})}
    $cf1 \gets \FLongestRef{T, k, previousFactors}$ \;
    $cf2 \gets \FLongestTruncation{T, $k + cf1.length$, previousFactors}$ \;
    \tcp{Use single letter for $F_{x_2}$ if no truncation found, or when factor has length $1$} \pushline
    \popline\nosemic \If{$k + cf1.length \leq n$ $\boldsymbol{and}$ $cf2.length \leq 1$}{
      $cf2 \gets \{ factor \gets T[k + cf1.length], \ length \gets 1 \}$ \;
    }
    \nosemic $cf \gets \{$ \;
      \pushline\nosemic $firstFactor \gets cf1.factor$, \;
      \nosemic $secondFactor \gets cf2.factor$, \;
      \nosemic $length \gets cf1.length + cf2.length$ \;
    \popline\nosemic $\}$ \;
    \tcp{Truncation factor (\Cref{alg:longesttruncation})}
    $tf \gets \FLongestTruncation{T, k, previousFactors}$ \;
    \Return longest factor from $cf$ and $tf$, where ties are broken such that combination is preferred \;
  }
\end{algorithm}

\block{\ref{it:longRef}}
We begin the traversal over the trie for computing the longest reference of $F_x$ (cf.~\cref{alg:longestref}).
As long as $\ibeg(F_x) + i \leq n$, we repeatedly execute the following steps:
First, we find the outgoing edge $e=(u,v)$ via $v = \child(u, T[\ibeg(F_x) + i])$, 
where the first letter of $e$'s label $e.\fnLabel[1]$ matches the current letter $T[\ibeg(F_x) + i]$.
If $v = \bot$, we stop the traversal.
Otherwise, we have to check that $e.\fnLabel$ matches with the letters read from the text before going to the next node~$v$.
We do so by the LCE query 
$T.\LimitedLCE(\ibeg(F_x) + i, e.\fnPos{}, e.\fnLen{})$
between $e.\fnLabel{}$ and $T[\ibeg(F_x) + i..]$, and check whether the returned length is at least $|e|$ (Line~\ref{alg:longestref:edgelce}).
If the returned length is smaller than $|e|$, we stop the traversal.
Otherwise, we increase the number of read letters $i$ by $|e|$, 
and since we have reached the next node~$v$, 
update the node pointer $u$ to $v$.
If $v$ is a factor node, we update $y \gets \factorIndex(u)$ (Line~\ref{alg:longestref:updatefactor}).
By doing so, upon finishing the traversal, $y$ stores the index of the longest reference.
If $|F_y| \ge 2$ we return $y$.
Otherwise, if the remaining input $T[\ibeg(F_x)..]$ is not empty and $|F_y| \le 1$, we instead return the first letter of the remaining input $T[\ibeg(F_x)]$.
This concludes the search for the longest reference.

\block{\ref{it:Truncation}}
The search for the longest truncation of a previous factor (cf.~\cref{alg:longesttruncation}) can be adapted from the search for the longest reference.
First, after the trie traversal is finished, we immediately return $y$ regardless of whether $|F_y| \le 1$.
Second, we update the reference candidate~$y$ not only if we have reached a new factor node, but also if we have reached a split node (Line~\ref{alg:longesttruncation:alwaysupdate})
by setting $y$ to its \succIndex{}.
Finally, we have to handle the case where we were not able to reach the next node because the edge label does not fully match the letters read from the input.
We still stop the traversal over the trie in that case, but before that, we increase $i$ by the returned length of the $\LimitedLCE$ query, i.e., the number of matching letters on that edge label.
After increasing $i$, we update $y$ to the \succIndex{} of the end node of the edge and report the truncation length~$i$ (Line~\ref{alg:longesttruncation:extraupdate}).

Finally, we can use the two helper functions to compute $F_x$ (cf.~\cref{alg:nextlzdpfactor}).
To find the longest truncation factor, we simply call the truncation helper function.
To find the longest combination factor, we need some extra logic.
The $F_{x_1}$ part of the combination factor can be found by calling the longest reference helper function.
The $F_{x_2}$ part of the combination factor is either a truncation of a previous factor, including $F_0$, or a single letter.
The truncation part can be handled by calling the truncation helper function where the remaining input starts directly after the already determined $F_{x_1}$ part.
Afterward, we simply check if the remaining input $T[\ibeg(F_x)+|F_{x_1}|..]$ that starts after the $F_{x_1}$ part is not empty and the length of the returned truncation part is smaller or equal to 1.
If that is the case, we update $F_{x_2} \gets T[\ibeg(F_x)+|F_{x_1}|]$ to the first letter after the $F_{x_1}$ part with a length of 1.
Finally, we compare the lengths of the combination and truncation factor, and return the longest factor of the two. This concludes the algorithm for computing $F_x$.
Since all traversals for $F_x$ stop at the first mismatch, the algorithm has read only $T[1..\iend(F_x)+1]$ when it outputs~$F_x$; it thus works online.

\block{Time complexity}
We show that the proposed LZD+ algorithm takes $O(n)$ expected time.
For that we first show that computing $F_x$ takes $O(\ell+1)$ expected time, given $\ell := |F_x|$ is the length of $F_x$, using the pseudocode as reference to make a number of observations.
We call the basic building block of our algorithm an \emph{iteration step}, which is either a letter-wise comparison or a traversal step.
The letter-wise comparison emerges from calls to $T.\LimitedLCE$. 
A traversal step is a $\child$ operation, which we compute during a trie traversal (cf.~the inner body of the radix trie loop starting at Line~\ref{alg:longesttruncation:whiledef} of \cref{alg:longesttruncation}) to move further downwards the trie.

	\begin{enumerate}[wide, labelwidth=!, labelindent=0pt]
  \item If the longest truncation factor has a length of $\ell_t$, the search for that factor took at most $\ell_t + 1$ traversal steps.
 Since at least one letter gets added to the factor per traversal step, unless the radix trie loop is terminated during that traversal step because there is no corresponding edge (Line~\ref{alg:longesttruncation:extraupdate}).
 Similarly, since we extend $F_x$ by one per matching letter pair, unless the step turns out to be a letter mismatch, we match $\ell_t$ pairs of letters during the letter-wise comparisons, and additionally find at most one mismatching letter pair in the whole traversal to compute~$F_x$.
 The number of iteration steps for the search of the longest truncation factor is therefore upper-bounded by $\underbrace{(\ell_t + 1)}_{\text{visited nodes}} + \underbrace{\ell_t}_{\text{matching letter pairs}} + \underbrace{1}_{\text{mismatching letter pair}} \in O(\ell_t+1)$.

\item The search for $F_{x_1}$ takes the same number of iteration steps as the search for the longest truncation factor, since the only difference is when and how the reference candidate~$y$ is updated --- the traversal over the radix trie remains the same (\Cref{alg:longesttruncation,alg:longestref}).

\item The search for $F_{x_2}$ follows the same mechanism as the search for the truncation factor, and therefore, if $F_{x_2}$ has a length of $\ell_{c_2}$ after truncation (i.e., $\ell_{c_2} = \ell_x - \ell_{c_1}$, where $\ell_x$ is the length of the combination factor after truncation and $\ell_{c_1} = |F_{x_1}|$), finding the index $x_2$ of $F_{x_2}$ takes $O(\ell_{c_2})$ iteration steps (\Cref{alg:nextlzdpfactor}). In case that $F_{x_2}$ is a letter, we find it in amortized constant time.

  \item Neither $\ell_t$ nor $\ell_{c_2}$ can be greater than $\ell$, the length of the longest factor of the two, because $\ell = \max( \ell_{c_1} + \ell_{c_2}, \ell_t )$.
 Therefore, the total number of iteration steps is bounded by $2 \cdot O(1+ \ell_t) + O(\ell_{c_2}) \leq 2 \cdot O(1+\ell) + O(\ell) = O(1+\ell)$.

  \item Apart from the LCE queries, every operation per node traversal takes $O(1)$ expected time.
 The time complexity of all $\LimitedLCE$ queries is bounded by the number of letter-wise comparisons.
 Therefore, the expected running time of the trie traversal is bounded by the sum of the traversal steps and the letter-wise comparisons, i.e., the number of iteration steps, and is thus $O(\ell)$.

\item Similarly, the statements before and after each traversal also take $O(1)$ time, as well as the extra logic for $F_{x_2}$ to account for single letters (\Cref{alg:nextlzdpfactor}).

  \item Finding the longest factor of the two can be accomplished with the help of one comparison, and thus also takes $O(1)$ time.

  \item The total expected running time of finding $F_x$ with $\ell := |F_x|$ is therefore $O(\ell)$.

\end{enumerate}

Now, we show that the LZD+ algorithm takes $O(n)$ expected time.
First, we enter a loop to process each factor~$F_x$ individually:
We search $F_x$'s reference in the radix trie and subsequently insert $F_x$ into the radix trie.
 By the analysis above, we find $F_x$'s reference in $O(\ell)$ expected time, if $\ell$ is its length.
 Inserting a string of length $\ell$ into a radix trie takes $O(\ell)$ expected running time.
 The length of all factors sums to $n$, which means when there are $z$ factors in total, $\sum_{x=1}^z |F_x| = n$, and therefore, the total expected running time is $\sum_{x=1}^z \left( 2 \cdot O(|F_x|) \right) = O(n)$.

Finally, we can de-randomize our algorithm.
Recall that the randomization is only needed for the $\child$ operation implemented by a lookup in a hash table.
By switching from the hash table to a deterministic data structure like a balanced search tree with $O(\lg \sigma)$ lookup time, 
we obtain $O(n \lg \sigma)$ worst-case total time.

\block{Space complexity}
In addition to having read-only access to $T$, our proposed algorithm needs only $O(z)$ working space.
Inserting a factor of length $\ell$ into the radix trie increases the space complexity of the radix trie by a constant for creating a factor node and at most one split node, 
which means that after inserting all factors, the radix trie has a space complexity of $O(z)$.
All other variables in our algorithm take $O(1)$ space, therefore the total space complexity of the LZD+ algorithm is $O(z)$ on top of the text.
We summarize the established complexities in the following theorem.

\begin{theorem}\label[theorem]{thmLZDPlus}
    We can compute LZD+ online in $O(n)$ expected time or $O(n \lg \sigma)$ worst-case time with $O(z)$ words of working space in addition to the read-only text, where $z$ is the number of LZD+ factors.
\end{theorem}

\block{Known lower bound on the LZD factors}\label{sec:lowerbound}
Badkobeh et al.\ have shown that for arbitrarily large $n$, there are strings $S_k$ of length $\Theta(n)$ for which the size of the grammars produced by LZD is larger than the size of the smallest grammar generating $S_k$ by a factor of $\Omega(n^{\frac{1}{3}})$~\cite[Theorem 1]{badkobeh17two}.
They showed that, when $k \geq 4$ is a power of two, then $n = \Theta(k^3)$, and the size of the grammar corresponding to the LZD factorization of $S_k$ is $\Omega(k^2)$ whereas the size of the smallest grammar is $O(k)$. Their proof analyzes only an analogous string for LZMW in detail; for LZD, they omit the analysis as similar and confirm the bound experimentally for small~$k$. \Cref{thm:binary} in \cref{sec:bounds} gives a complete analysis for a binary variant of these strings.
In detail, $S_k$ has the following shape:

\begin{flalign*}
& S_k = \left( \texttt{a}^2 \texttt{c}^2 \texttt{a}^3 \texttt{c}^3 \cdots \texttt{a}^k \texttt{c}^k \right) \left( \texttt{bb} \texttt{abb} \texttt{a}^2 \texttt{bb} \texttt{a}^3 \cdots \texttt{bb} \texttt{a}^{k-1} \texttt{bb} \right) \left( \Delta_0 \texttt{d}^2 \Delta_1 \texttt{d}^3 \cdots \Delta_k \texttt{d}^{k+2} \right) V_k^{\frac{k}{2}} \text{~where} && \\
& \Delta_i = \texttt{a}^i \texttt{bb} \texttt{a}^{k-i} \text{~and~}
V_k = \Delta_k \Delta_{k-1} \Delta_k \Delta_{k-2} \Delta_k \Delta_{k-3} \cdots \Delta_k \Delta_{k/2 + 1} \Delta_k \texttt{a}^{k-1} && 
\end{flalign*}

Intuitively, $S_k$ consists of a \emph{training prefix}~$P_k$, the part before $V_k^{k/2}$, and a periodic \emph{hard suffix}~$V_k^{k/2}$, i.e., $k/2$ copies of the \emph{hard part}~$V_k$.
The training prefix fills the dictionary with runs of~$\texttt{a}$, the strings $\texttt{a}^i\texttt{bb}$, and the strings~$\Delta_i$, which all have length $k+2$ and differ only in the position of~$\texttt{bb}$; the runs of $\texttt{c}$ and~$\texttt{d}$ merely separate these blocks.
LZD parses each copy of the hard part~$V_k$ into $k/2$ factors that cover $|V_k|+1$ letters, so the next copy is entered one letter later, where the factors of earlier copies are out of phase and do not match; this costs $\Theta(k)$ new factors per copy and $\Theta(k^2)$ in total.
A grammar, in contrast, needs only $O(k)$ rules: it builds the runs $\texttt{a}^i$, $\texttt{c}^i$, and~$\texttt{d}^i$ incrementally, each~$\Delta_i$ with $O(1)$ further rules, $V_k$ with $O(k)$ rules, and $V_k^{k/2}$ with $O(\log k)$ rules by repeated doubling.

 \begin{table}[t]
	\begin{minipage}{0.8\linewidth}
  \centering
  \begin{tabular}{|l|r|r|r|r|r|r|r|}
    \hline
    & $k=4$ & $k=8$ & $k=16$ & $k=32$ & $k=64$ & $k=128$ & $k=256$ \\\hline
    LZD & 24 & 56 & 144 & 416 & 1344 & 4736 & 17664 \\\hline
    LZD+ & 24 & 56 & 144 & 416 & 1344 & 4736 & 17664 \\\hline
    LZDR & 24 & 51 & 99 & 195 & 387 & 771 & 1539 \\\hline
    $6k + 3$ & 27 & 51 & 99 & 195 & 387 & 771 & 1539 \\\hline
  \end{tabular}
	\end{minipage}
	\begin{minipage}{0.17\linewidth}
		\caption{Factor count of $S_k$ described at the end of \cref{sec:lzd}}
	\label{tab:lowerBound}
	\end{minipage}
\end{table}

In \cref{tab:lowerBound}, we practically evaluated whether 
LZD+ and LZDR exhibit the same bound for the provided string $S_k$,
where LZDR is a compression scheme that we will introduce in the next section.
On the one hand, we observe that the factor count of LZD+ for $S_k$ matches that of LZD up to $k=256$.
Regardless of whether this pattern continues, LZD+ does not improve on the bound: \cref{thm:binary} in \cref{sec:bounds} gives strings on which LZD+ produces $\Omega(n^{2/3})$ factors, whereas a grammar of size $O(n^{1/3})$ exists.
On the other hand, LZDR appears to have a much lower factor count, and the factor count matches the linear expression $6k + 3$ for $k \in \{2^3, 2^4, \ldots, 2^8\}$.
We do not pursue this observation further, since it does not extend to other strings: \cref{thm:binary} shows that a rotation of the hard part already forces LZDR to produce $\Omega(n^{2/3})$ factors.

\section{LZDR compression scheme}
We propose a greedy compression scheme, called LZDR, that is based on LZD+ with one modification: we replace the truncation production rule introduced in LZD+ with a production rule that finds the longest repetition of a previous factor or of a single letter, which can also be truncated.
Although LZD+ and LZDR both have a linear-time compression algorithm, LZDR has the advantage that it seems to avoid the bound from Badkobeh et al.\ on their provided example string, as highlighted in the previous section (\cref{tab:lowerBound}); \cref{thm:binary} shows, however, that this advantage does not extend to all strings.

\subsection{Definition of LZDR}
The LZDR (LZD-runlength compressed) factorization of $T$ is the factorization $T = F_1 F_2 \cdots F_z$ such that, for every $x \in [1..z]$, the two possible production rules for $F_x$ are:
\begin{enumerate}
    \item Combination: $p_1 = \left( F_{x_1} F_{x_2} \right) [1..\ell_x]$ where $p_1$ is a prefix of $T[\ibeg(F_x)..n]$ and $F_{x_1}$ is the longest prefix of $T[\ibeg(F_x)..n]$ with $F_{x_1} \in \{ F_1, \ldots, F_{x-1} \} \cup \Sigma$, $F_{x_2} \in \{ F_0, F_1, \ldots, F_{x-1} \} \cup \Sigma$, and $1 \leq \ell_x \leq |F_{x_1}| + |F_{x_2}|$.
    \item Repetition: $p_2 = \left( F_y^\infty \right) [1..\ell_x]$ where $p_2$ is a prefix of $T[\ibeg(F_x)..n]$ with $F_y \in \{ F_1, \ldots, F_{x-1} \} \cup \Sigma$, and $\ell_x \geq 2$.
\end{enumerate}
The factor $F_x$ is then chosen among all possible $p_1$'s and $p_2$'s as the one that gives the maximum length. 
If $F_{x_1}$ or $F_{x_2}$ have a length of 1, we prefer to use a single letter instead of a previous factor.

In the remainder of this paper we differentiate between two cases regarding the repetition rule.
We speak of \textit{truncation} if $\ell_x \leq |F_y|$.
Otherwise, if $\ell_x > |F_y|$, we will speak of \textit{repetition}.
With the help of the repetition production rule, special string families like $\texttt{a}^n$ with $n \in \mathbb{N}^+$ are parsed to a single factor in LZDR\@.
In contrast, LZD parses a string of the form $\texttt{a}^n$ with $n = 2^{k+1} - 2$ and $k \in \mathbb{N}^+$ to $k$ factors since it can only make use of its combination rule.
This means, string families exist with $\Theta(\log n)$ LZD factors and $\Theta(1)$ LZDR factors.

Intuitively, $(F_y^\infty)[1..\ell_x]$ is the prefix of length~$\ell_x$ of the infinite string $F_yF_yF_y\cdots$, i.e., about $\ell_x/|F_y|$ copies of~$F_y$, the last one possibly truncated.
For $\ell_x \le |F_y|$, the rule is the truncation rule of LZD+, so the dictionary of LZDR is prefix-closed as well.
Since a letter is always an admissible reference, every factor of LZD+ and of LZDR except possibly the last one has length at least two; hence, both factorizations have at most $\lceil n/2 \rceil$ factors.

\begin{table}
  \scriptsize
  \centering
  \begin{tabular}{|c|c|c|c|c|c|c|c|c|c|}
    \hline
    $\boldsymbol{F_1}$ & $\boldsymbol{F_2}$ & $\boldsymbol{F_3}$ & $\boldsymbol{F_4}$ & $\boldsymbol{F_5}$ & $\boldsymbol{F_6}$ & $\boldsymbol{F_7}$ & $\boldsymbol{F_8}$ & $\boldsymbol{F_9}$ & $\boldsymbol{F_{10}}$ \\\hline
    aa & bb & aabb & baabb & bbba & ba & baab & cccc & babab & c \\\hline
    $(\texttt{a}, \texttt{a})$ & $(\texttt{b}, \texttt{b})$ & $(F_1, F_2)$ & $(\texttt{b}, F_3)$ & $(F_2, F_4)[1..4]$ & $(\texttt{b}, \texttt{a})$ & $(F_4)^1 [1..4]$ & $\texttt{c}^4$ & $(F_6)^3 [1..5]$ & $(\texttt{c}, F_0)$ \\\hline
  \end{tabular}
  \caption{LZDR factorization of the string \texttt{aabbaabbbaabbbbbababaabccccbababc}.
In practice, we can represent the factors as 
$F_1 = (\texttt{a}, \texttt{a})$,
$F_2 = (\texttt{b}, \texttt{b})$,
$F_3 = (F_1, F_2)$,
$F_4 = (\texttt{b}, F_3)$,
$F_5 = (F_2, F_4)[1..4]$,
$F_6 = (\texttt{b}, \texttt{a})$,
$F_7 = (F_4)^1 [1..4]$,
$F_8 = \texttt{c}^4$,
$F_9 = (F_6)^3 [1..5]$,
and $F_{10} = (\texttt{c}, F_0)$.
  }
	\label{tab:lzdrEx}
\end{table}

\begin{example}\label[example]{ex:lzdrEx}
The LZDR factorization of the string \texttt{aabbaabbbaabbbbbababaabccccbababc} is $F_1 = \texttt{aa}$ (Combination of $\texttt{a}$ and $\texttt{a}$),
$F_2 = \texttt{bb}$ (Combination of $\texttt{b}$ and $\texttt{b}$),
$F_3 = \texttt{aabb}$ (Combination of $F_1$ and $F_2$),
$F_4 = \texttt{baabb}$ (Combination of $\texttt{b}$ and $F_3$),
$F_5 = \texttt{bbba}$ (Combination of $F_2$ and $F_4$ with truncation applied),
$F_6 = \texttt{ba}$ (Combination of $\texttt{b}$ and $\texttt{a}$),
$F_7 = \texttt{baab}$ (Truncation of $F_4$ via repetition rule),
$F_8 = \texttt{cccc}$ (Repetition of $\texttt{c}$),
$F_9 = \texttt{babab}$ (Repetition of $F_6$),
and $F_{10} = \texttt{c}$ (Combination of $\texttt{c}$ and $F_0$).
We visualized the factorization in \cref{tab:lzdrEx} and give the radix trie in \cref{fig:radixtrieLZDR}.
\end{example}

\begin{figure}
  \hspace{0.05\textwidth}
  \begin{subfigure}[t]{0.18\textwidth}
    \centering
    \begin{forest}
      for tree={circle,draw, l sep=20pt, minimum size=32pt, inner sep=0pt}
      [0(0)
        [1(1),double,edge label={node[midway,left] {aa}} ]
      ]
    \end{forest}
    \caption{After insertion of $F_1 = \texttt{aa}$}
    \label{fig:radixtrieLZDRA}
  \end{subfigure}
  \hspace{0.05\textwidth}
  \begin{subfigure}[t]{0.18\textwidth}
      \centering
      \begin{forest}
        for tree={circle,draw, l sep=20pt, minimum size=32pt, inner sep=0pt}
        [0(0)
          [1(1),double,edge label={node[midway,left] {aa}} ]
          [2(2),double,edge label={node[midway,right] {bb}} ]
        ]
      \end{forest}
      \caption{After insertion of $F_2 = \texttt{bb}$}
      \label{fig:radixtrieLZDRB}
  \end{subfigure}
  \hspace{0.05\textwidth}
  \begin{subfigure}[t]{0.40\textwidth}
      \centering
      \begin{forest}
        for tree={circle,draw, l sep=20pt, minimum size=32pt, inner sep=0pt}
        [0(0)
          [1(1),double,edge label={node[midway,left] {aa}}, tier=d1
            [3(3),double,edge label={node[midway,left] {bb}}, tier=d2 ]
          ]
          [0(2),edge label={node[midway,right] {b}}, tier=d1
            [6(6),double,edge label={node[midway,left] {a}}, tier=d2
              [7(7),double,edge label={node[midway,left] {ab}}, tier=d3
                [4(4),double,edge label={node[midway,left] {b}}, tier=d4 ]
              ]
              [9(9),double,edge label={node[midway,right] {bab}}, tier=d3 ]
            ]
            [2(2),double,edge label={node[midway,right] {b}}, tier=d2
              [5(5),double,edge label={node[midway,right] {ba}}, tier=d3 ]
            ]
          ]
          [10(10),font=\scriptsize,double,edge label={node[midway,right] {c}}, tier=d1
            [8(8),double,edge label={node[midway,right] {ccc}}, tier=d2 ]
          ]
        ]
      \end{forest}
      \caption{After all factors from $F_1$ to $F_{10}$ have been inserted}
      \label{fig:radixtrieLZDRC}
  \end{subfigure}
  \caption{Radix trie for LZDR factorization of \texttt{aabbaabbbaabbbbbababaabccccbababc}. Double circled nodes represent factor nodes, while single circled nodes represent split nodes. The first number represents the index of the factor node, and the number in parentheses is the \succIndex{}.}
  \label{fig:radixtrieLZDR}
\end{figure}

\begin{algorithm}[t]
  \scriptsize
  \DontPrintSemicolon
  \caption{Finding the next longest repetition of a previous factor or single letter.
We assume that $\Sigma$ and the range of factor indices $[0..z]$ are disjoint such that 
we can interpret the output correctly even if the returned reference is a letter run in Line~\ref{alg:longestrepetition:letterRepetition}, or a reference to a previous factor.
For instance, a pragmatic way is to encode letters by negative numbers.
  }
  \label{alg:longestrepetition}
  \KwData{Entire input string $T$ of length $n$, starting position for current factor $k \in \mathbb{N}^+$, $previousFactors$ as radix trie}
  \SetKwProg{Def}{def}{:}{}
  \SetKwFunction{FLongestRepetition}{GetLongestFactorRepetition}
  \Def{\FLongestRepetition{$T$, $k$, previousFactors}}{
    \tcp{Initialize the longest factor with the empty factor $F_0$}
    $y \gets \{ factor \gets 0, \ length \gets 0 \}$ \;
    \tcp{Initialize the current node and current index}
    $u \gets \text{root node of $previousFactors$}$ \;
    $i \gets 0$ \;
    \While{$k + i \leq n$}{
      \tcp{Get end node of edge that starts with the current letter}
      $v \gets u.\child(T[k + i])$ \;
      \If{$v = \bot$}{
        \tcp{Edge does not exist}
        \tcp{We were unable to read the letter, therefore break loop}
        \Break \;
      }
      \tcp{Get edge from $u$ to $v$}
      $e \gets (u, v)$ \;

      \tcp{Find length of longest common prefix of $T[k+i..]$ and edge label of $e$}
      $commonLength \gets T.\LimitedLCE(k + i, e.\fnPos, e.\fnLen)$ \;

      \eIf{$commonLength \geq e.\fnLen$}{
        \tcp{edge label successfully read, increase $i$}
        $i \gets i + e.\fnLen$ \;
        \tcp{Go to next node}
        $u \gets v$ \;
        \If{$\text{$u$ is not split node}$}{
          \tcp{Update factor if longer}
          $repetitionLength \gets i + T.\fnLCE(k, k + i)$ \; \label{alg:longestrepetition:changedupdate}
          \If{$repetitionLength > y.length$}{
            $y \gets \{ factor \gets \factorIndex(u), \ length \gets repetitionLength \}$ \;
	    \label{alg:longestrepetition:letterRepetition}
          }
        }
      }{
        \tcp{Mismatch during edge label of $e$, could not reach next node}
        \Break \;
      }
    }

    \tcp{Use single letter repetition if longer or equal to current length}
    \If{$k \leq n$}{
      $repetitionLength \gets 1 + T.\fnLCE(k, k + 1)$ \; \label{alg:longestrepetition:singleupdate}
      \If{$repetitionLength \geq y.length$}{
        $y \gets \{ factor \gets T[k], \ length \gets repetitionLength \}$ \;
      }
    }

    \Return $y$ \;
  }
\end{algorithm}

\begin{algorithm}[t]
  \scriptsize
  \DontPrintSemicolon
  \caption{Finding the next longest LZDR factor in $O(\ell)$ time}
  \label{alg:nextlzdrfactor}
  \KwData{Entire input string $T$ of length $n$, starting position for current factor $k \in \mathbb{N}^+$, $previousFactors$ as radix trie}
  \KwResult{the next longest LZDR factor of length $\ell$}
  \SetKwProg{Def}{def}{:}{}
  \SetKwFunction{FLongestLZDRFactor}{NextLongestLzdrFactor}
  \SetKwFunction{FLongestRef}{GetLongestFactorRef}
  \SetKwFunction{FLongestTruncation}{GetLongestFactorTruncation}
  \SetKwFunction{FLongestRepetition}{GetLongestFactorRepetition}
  \Def{\FLongestLZDRFactor{$T$, $k$, previousFactors}}{
    \tcp{Combination factor (\Cref{alg:longestref,alg:longesttruncation})}
    $cf1 \gets \FLongestRef{T, k, previousFactors}$ \;
    $cf2 \gets \FLongestTruncation{T, $k + cf1.length$, previousFactors}$ \;
    \tcp{Use single letter for $F_{x_2}$ if no truncation found, or when factor has length $1$} \pushline
    \popline\nosemic \If{$k + cf1.length \leq n$ $\boldsymbol{and}$ $cf2.length \leq 1$}{
      $cf2 \gets \{ factor \gets T[k + cf1.length], \ length \gets 1 \}$ \;
    }
    \nosemic $cf \gets \{$ \;
      \pushline\nosemic $firstFactor \gets cf1.factor$, \;
      \nosemic $secondFactor \gets cf2.factor$, \;
      \nosemic $length \gets cf1.length + cf2.length$ \;
    \popline\nosemic $\}$ \;
    \tcp{Truncation factor (\Cref{alg:longesttruncation})}
    $tf \gets \FLongestTruncation{T, k, previousFactors}$ \;
    \tcp{Repetition factor (\Cref{alg:longestrepetition})}
    $rf \gets \FLongestRepetition{T, k, previousFactors}$ \;
    \Return longest factor from $cf$, $tf$ and $rf$, where ties are broken such that combination is preferred most and repetition least \;
  }
\end{algorithm}

\subsection{Linear-time compression algorithm for LZDR}

We propose a linear-time offline algorithm for LZDR\@.
For that, we preprocess the text. 
In detail, we build an LCE data structure on $T$ in $O(n)$ time that will allow us to answer $T.\fnLCE$ in $O(1)$ time~\cite{ilie10longest}.
In the actual factorization, we repeatedly compute the next factor~$F_x$ that is a prefix of the remaining input~$T[\ibeg(F_x)..]$.
For that, we describe a new helper function for computing $F_x$ in addition to the ones from the LZD+ algorithm. 
We show how we can adapt the search for the longest reference (\Cref{alg:longestref}) to find the next longest repetition of a previous factor or of a single letter (\Cref{alg:longestrepetition}).
To find the longest repetition of a previous factor instead of the longest previous factor itself, we modify how the reference candidate~$y$ is updated during the radix trie traversal.

Suppose that we want to compute the factor~$F_x$.
Up until now, the reference length was determined by the length of reference candidate $y$, i.e., $|F_y|$. 
Having the possibility to also use repetitions, we have to maintain the reference length~$\ell_x$ separately.
Another difference is that before, whenever a factor node was reached, the reference candidate $y$ got updated to that factor.
Now, instead of updating~$y$, we first determine the repetition length of the reached node via $i + T.\fnLCE(\ibeg(F_x), \ibeg(F_x) + i)$ (Line~\ref{alg:longestrepetition:changedupdate}). This is the length of the longest prefix of $T[\ibeg(F_x)..]$ with period~$i$: the text agrees with the repetition of the reached factor during its first $i$ letters, and afterwards exactly as long as it agrees with itself shifted by $i$ letters.
If the repetition length is greater than the reference length, we update the reference candidate $y$ to the index of the newly reached node and update the reference length~$\ell_x$.
The third modification is after the trie traversal: 
since we also allow for single letter repetitions, we determine the longest single letter repetition via $1 + T.\fnLCE(\ibeg(F_x), \ibeg(F_x) + 1)$ if the remaining input is not empty (Line~\ref{alg:longestrepetition:singleupdate}).
If that length is at least the length of the reference candidate, we update the reference candidate to the first letter of the remaining input with a corresponding length of the calculated repetition length.
Finally, we return the reference candidate $y$.

Now we can use the helper functions to compute~$F_x$ (\Cref{alg:nextlzdrfactor}).
To find the longest truncation and repetition factor, we simply call the truncation and repetition helper function.
To find the longest combination factor, we proceed as described in our LZD+ algorithm.
Finally, we compare the lengths of the combination factor, truncation factor, and repetition factor, and return the longest factor of the three. 

\block{Time complexity}
We show that the proposed LZDR algorithm takes $O(n)$ expected time.
For that, we base our analysis on the results for LZD+ and focus on the two differences to the LZD+ algorithm.
First, before we start searching for any factors, we build an LCE data structure upon $T$ in $O(n)$ time, which enables us to answer LCE queries in $O(1)$ time~\cite{ilie10longest}.
Second, we also consider the longest repetition of a previous factor or single letter in our search for the next factor (\Cref{alg:nextlzdrfactor}).
This search takes the same number of traversal steps as the search for the longest truncation factor, as the only difference is when and how the longest factor is updated --- the traversal over the trie stays the same (\Cref{alg:longestrepetition,alg:longesttruncation}).
Since we are now able to answer LCE queries in $O(1)$ time, and all other statements in that loop for the search of the longest repetition factor also take $O(1)$ time or $O(1)$ expected time, the search has the same bound as the search for the truncation factor.
That is, finding the next longest repetition factor, and by extension LZDR factor, takes $O(\ell)$ expected time for $\ell := |F_x|$.
These two differences do not change the time complexity of the algorithm, and therefore, the total expected running time of the LZDR algorithm is $O(n)$.

While we showed an expected running time of $O(\ell)$ for finding the next factor, a more precise bound is $O(\nu)$ expected time, where $\nu \le \ell + 1$ is the number of visited nodes in the radix trie.
This is due to the fact that with our LCE data structure, we can answer $T.\LimitedLCE$ in $O(1)$ time and thus can traverse an edge in $O(1)$ expected time.

\block{Space complexity}
In comparison to our LZD+ algorithm, which takes $O(z)$ space additional to $T$, the LZDR algorithm builds an LCE data structure that takes $O(n)$ space.
Therefore, the total space complexity of the LZDR algorithm is $O(n)$.
Finally, we can switch analogously to our LZD+ algorithm the hash table with a deterministic data structure, to obtain the following summary of the complexities for LZDR.

\begin{theorem}\label[theorem]{thmLZDR}
    We can compute LZDR in $O(n)$ expected time or $O(n \lg \sigma)$ worst-case time with $O(n)$ words of working space.
\end{theorem}

\subsection{Flexible parsing variants}\label{sec:flexible}
In contrast to greedy parsing, flexible parsing is semi-greedy with a one-step lookahead~\cite{matias99optimality}.
Flexible parsing empirically performs better than greedy parsing and achieves a lower factor count~\cite{horspool95effect,matias01effect} if the used dictionaries are prefix-closed.
We now define three different flexible parsing variants for LZDR,
where the first also has the theoretical guarantee that the factor count of the flexible parsing is at most the factor count of the greedy parsing (\cref{prop:flex}).

\block{Standard Flexible LZDR}
The \emph{Standard Flexible} LZDR (\StdFlex{}-LZDR) variant works similar to the flexible parsing variant introduced by Matias et al.~\cite{matias99optimality}.
This variant can only reference the previous factors computed by the \emph{greedy} LZDR parsing.
That means the following.
Suppose that the greedy LZDR parsing of $T$ is $T = R_1 R_2 \cdots R_{\zeta}$, such that $R_x$ is a LZDR factor, for each $x \in [1..\zeta]$.
The \StdFlex{}-LZDR parsing of $T$ is then $T = F_1 F_2 \cdots F_{z}$ such that, for every $x \in [1..{z}]$, $F_x$ is chosen as the non-greedy factor using the production rules of LZDR that maximizes the combined length $|F_x G_{x+1}|$ of $F_x$ and the lookahead factor~$G_{x+1}$, where we break ties in favor of a longer $F_x$ factor in all flexible parsing variants.
Here, $G_{x+1}$ is the lookahead factor that we choose greedily as the next longest factor, which is not necessarily equal to $F_{x+1}$ because we determine $F_{x+1}$ similarly in a semi-greedy way with a one-step lookahead after having processed~$F_x$.
The factor $F_x$ can only reference factors $R_y \in \{ F_0, R_1, R_2, \ldots, R_{\zeta} \}$ with $\iend(R_y) < \ibeg(F_x)$, that means it can only reference factors from the original LZDR parsing that end before the factor $F_x$ starts. (Otherwise, decompression is impossible in general.)
In the same sense, the lookahead factor $G_{x+1}$ can reference factors $R_y \in \{ F_0, R_1, R_2, \ldots, R_{\zeta} \}$ with $\iend(R_y) < \ibeg(F_x) + |F_x|$.

The optimality theorem of Matias et al.~\cite{matias99optimality} does not apply verbatim, since the phrases of LZDR are not the elements of a fixed prefix-closed dictionary: the first component of a combination is pinned to a longest complete reference, and the references available to \StdFlex{}-LZDR depend on the current position. We therefore prove the guarantee directly in \cref{prop:flex}.

To state the guarantee, we view a factorization as a path through the positions $0, 1, \ldots, n$ between the letters of~$T$: position~$p$ is the boundary directly after $T[p]$, i.e., the first $p$ letters have been parsed, and a factor $T[p+1..q]$ moves the parsing from position~$p$ to position~$q$.
Recall that $T = R_1 R_2 \cdots R_\zeta$ is the greedy LZDR factorization.
A \StdFlex{}-LZDR factor starting at $T[p+1]$ may reference the letters, the empty factor~$F_0$, and exactly those greedy factors that lie completely within the parsed prefix $T[1..p]$.
Since the greedy factors are consecutive, these references form the \emph{dictionary at position~$p$},
\[
  D(p) := \Sigma \cup \{F_0, R_1, R_2, \ldots, R_y\},
  \text{where $y$ is the largest index with $\iend(R_y) \le p$}
\]
(and $y = 0$ if $p < \iend(R_1)$).
Put in words, $D(p)$ grows only at the end positions of greedy factors, and it depends only on~$p$, not on how the flexible parsing has reached~$p$; the factors chosen by \StdFlex{}-LZDR itself never enter it.

Let the \emph{reach} $M(p)$ of a position $p < n$ be the largest end position of a factor that starts at $T[p+1]$ and is formed by the production rules of LZDR with references from~$D(p)$, i.e., $p$ plus the length of the longest such candidate; we set $M(n) := n$.
In words, $M(p)$ is how far a single factor can reach from position~$p$.
As for LZD+, the combination rule realizes every length in $[1..|F_{x_1}|+|F_{x_2}|]$ and the repetition rule every length from~$2$ up to its maximum, so the possible end positions of a factor starting at $T[p+1]$ form the whole interval $[p+1..M(p)]$.
For a factor~$F_x$ starting at $T[p+1]$ and ending at~$q$, the lookahead factor~$G_{x+1}$ is the longest factor starting at $T[q+1]$ with references from~$D(q)$, so $|F_xG_{x+1}| = M(q) - p$.
Hence, \StdFlex{}-LZDR moves from position~$p$ to the end position $q \in [p+1..M(p)]$ that maximizes~$M(q)$, and to the largest such~$q$ in case of ties, whereas the greedy parsing always moves to $M(p)$ itself.

\begin{figure}[t]

\centering
\begin{tikzpicture}[x=0.62cm,y=0.3cm,font=\scriptsize]
\foreach \s/\e/\i in {0/3/1,3/5/2,5/10/3,10/12/4,12/16/5,16/17/6} {
    \draw[|-|,thin] (\s+0.06,0) -- (\e-0.06,0);
    \node[above] at ({(\s+\e)/2},0.1) {$R_{\i}$};
  }
\foreach \c [count=\i] in {a,a,a,b,a,b,a,a,a,a,a,a,b,a,a,a,b}
    \node at (\i-0.5,-1.6) {\texttt{\c}};
  \foreach \p in {0,...,17} {
    \draw[gray!70] (\p,-0.8) -- (\p,-2.4);
    \node at (\p,-3.2) {$\p$};
  }
  \node[anchor=east] at (-0.7,-3.2) {position $p$};
\foreach \p/\y in {0/0,1/0,2/0,3/1,4/1,5/2,6/2,7/2,8/2,9/2,10/3,11/3,12/4,13/4,14/4,15/4,16/5}
    \node at (\p,-4.6) {$\y$};
  \node[anchor=east] at (-0.7,-4.6) {$y$ in $D(p)$};
\foreach \r/\t in {3/1,5/2,10/3,14/4,17/5} {
    \draw[densely dashed,gray] (\r,-5.6) -- (\r,-23.4);
    \node at (\r,-24.3) {$\rho(\t)$};
  }
\node[anchor=east] at (-0.7,-6.4) {reach $M(p)$};
  \node at (-1.3,-7.6) {G};
  \node at (-0.6,-7.6) {F};
\foreach \p/\x in {0/-1.12,3/-1.12,5/-1.12,9/-0.45,10/-1.12,12/-1.12,14/-0.45,16/-1.12}
    \draw[black!50,line width=0.9pt,line cap=round,dash pattern=on 0pt off 3.2pt] (\x,{-8.6-0.85*\p}) -- ({\p-0.12},{-8.6-0.85*\p});
  \foreach \p/\m in {0/3,1/3,2/4,3/5,4/6,5/10,6/12,7/12,8/12,9/14,10/12,11/16,12/16,13/17,14/17,15/17,16/17} {
    \draw[black!60,thick] (\p,{-8.6-0.85*\p}) -- (\m,{-8.6-0.85*\p});
    \draw[black!60] (\m,{-8.6-0.85*\p+0.3}) -- (\m,{-8.6-0.85*\p-0.3});
    \fill[black!60] (\p,{-8.6-0.85*\p}) circle (1.3pt);
  }
  \foreach \p in {0,3,5,10,12,16} \node at (-1.3,{-8.6-0.85*\p}) {$\circ$};
  \foreach \p/\q in {0/3,3/5,5/9,9/14,14/17} {
    \draw[orange!85!black,very thick] (\p,{-8.6-0.85*\p}) -- (\q,{-8.6-0.85*\p});
    \fill[orange!85!black] (\q,{-8.6-0.85*\p}) circle (2pt);
    \node at (-0.6,{-8.6-0.85*\p}) {$\bullet$};
  }
\end{tikzpicture}
\caption{Notions of \cref{prop:flex} for the string of \cref{tab:flexEx}.
Top: the greedy LZDR factors $R_1, \ldots, R_6$ above the text, the positions~$p$, and the number~$y$ of greedy factors in $D(p) = \Sigma \cup \{F_0, R_1, \ldots, R_y\}$.
Bottom: for each position~$p$, a bar from $p$ to its reach $M(p)$ shows the end positions of all factors starting at $T[p+1]$.
The greedy parsing (positions marked in column~G) always moves to the end of the bar; \StdFlex{}-LZDR (column~F, orange) moves to the dot, the end position~$q$ with the largest reach $M(q)$.
The dashed lines mark the frontiers $\rho(1), \ldots, \rho(5)$, the farthest positions reachable with $1, \ldots, 5$ factors.}
\label{fig:flex}

\end{figure}

\Cref{fig:flex} illustrates these notions on the string of \cref{tab:flexEx}, whose greedy LZDR factorization is $R_1 \cdots R_6 = \texttt{aaa} \cdot \texttt{ba} \cdot \texttt{baaaa} \cdot \texttt{aa} \cdot \texttt{baaa} \cdot \texttt{b}$.
For instance, $D(9) = \Sigma \cup \{F_0, R_1, R_2\}$, because $R_3$ ends only at position~$10$; the factor \texttt{baaa} of \StdFlex{}-LZDR ending at position~$9$ is not added.
At position~$5$, the greedy parsing takes the longest factor $R_3 = \texttt{baaaa}$ and moves to $M(5) = 10$, but from there, a single factor reaches only $M(10) = 12$.
\StdFlex{}-LZDR instead compares the reaches $M(6), \ldots, M(10) = 12, 12, 12, 14, 12$ and moves to position~$9$, from which the factor $\texttt{aaaba} = R_1R_2$ reaches position~$14$.
In total, it needs five factors, one fewer than the greedy parsing.

\begin{theorem}\label[theorem]{prop:flex}

For every text~$T$, \StdFlex{}-LZDR produces the minimum number of factors among all factorizations of~$T$ in which every factor starting at $T[p+1]$ ends at a position in $[p+1..M(p)]$.
In particular, it produces at most as many factors as the greedy LZDR factorization.

\end{theorem}
\begin{proof}

Let the \emph{frontier} $\rho(t)$ be the largest position reachable from~$0$ by at most~$t$ such factors, with $\rho(0) = 0$; in \cref{fig:flex}, the frontiers are the dashed lines.
By induction on~$t$ and the interval property, every position in $[0..\rho(t)]$ is reachable by at most $t$ factors, so the next frontier is the largest reach of these positions:
\[
  \rho(t+1) = \max_{0 \le q \le \rho(t)} M(q).
\]
Since $M(q) \ge q+1$ for $q < n$, we have $\rho(t+1) > \rho(t)$ whenever $\rho(t) < n$.

Let $f_0 := 0$, and let $f_t$ be the end position of the $t$th factor of \StdFlex{}-LZDR.
We show by induction on~$t$ that $M(f_t) = \rho(t+1)$, i.e., that after $t$ factors, \StdFlex{}-LZDR stands at a position from which one more factor reaches the next frontier; for $t = 0$, this is $\rho(1) = M(0)$.
Assume that $M(f_t) = \rho(t+1) < n$.
Since $f_t \le \rho(t)$, every $q \le f_t$ satisfies $M(q) \le \rho(t+1) < \rho(t+2)$, so every maximizer~$q$ in $\rho(t+2) = \max_{0 \le q \le \rho(t+1)} M(q)$ lies in $[f_t+1..\rho(t+1)] = [f_t+1..M(f_t)]$.
This is exactly the range from which \StdFlex{}-LZDR chooses~$f_{t+1}$ by maximizing $M(f_{t+1})$; hence, $M(f_{t+1}) = \rho(t+2)$.
Consequently, \StdFlex{}-LZDR reaches position~$n$ with the smallest number~$t$ of factors for which $\rho(t) = n$.
In \cref{fig:flex}, $f_t = 0, 3, 5, 9, 14$ and $M(f_t) = 3, 5, 10, 14, 17 = \rho(t+1)$ for $t = 0, \ldots, 4$, whereas the greedy parsing falls behind at position~$10$, since $M(10) = 12 < \rho(4)$.

Finally, the greedy LZDR factorization is one of the considered factorizations: at the starting position~$p$ of a greedy factor, $D(p)$ consists of the letters and exactly those greedy factors computed so far, so the greedy factor ends at~$M(p)$.

\end{proof}

\begin{table}[t]
  \scriptsize
	\hspace{-5em}
  \begin{tabular}{|l|r|r|r|r|r|r|r|r|r|}
        \hline
		\multicolumn{10}{|c|}{Calgary corpus}\\
        \hline
    {} &
		\multicolumn{4}{c|}{LZDR}&
    {} &
    {} &
		\multicolumn{3}{c|}{LZW}
		\\
		{dataset} &
    {greedy} &
		{\StdFlex{} } &
		{\AltFlex{} } &
		{\MaxAlt{}} &
    {LZD+} &
    {LZ78} &
    {greedy} &
		{\StdFlex{} } &
		{\AltFlex{} }
        \\\hline
  bib    & $-3.49\%$ & $\mathbf{-7.86\%}$ & $-7.03\%$ & $-5.84\%$ & $-2.83\%$ & $62.24\%$ & $103.08\%$ & $92.87\%$ & $84.65\%$ \\\hline
  book1  & $-3.43\%$ & $\mathbf{-8.37\%}$ & $-7.42\%$ & $-6.36\%$ & $-3.50\%$ & $53.87\%$ & $81.05\%$ & $75.94\%$ & $73.96\%$ \\\hline
  book2  & $-2.92\%$ & $\mathbf{-8.79\%}$ & $-8.11\%$ & $-6.33\%$ & $-2.74\%$ & $65.36\%$ & $94.67\%$ & $86.00\%$ & $80.79\%$ \\\hline
  geo    & $-1.43\%$ & $\mathbf{-2.16\%}$ & $-1.87\%$ & $-1.85\%$ & $-1.37\%$ & $12.04\%$ & $82.30\%$ & $79.65\%$ & $79.51\%$ \\\hline
  news   & $-3.40\%$ & $\mathbf{-7.40\%}$ & $-6.83\%$ & $-5.57\%$ & $-3.49\%$ & $56.82\%$ & $94.13\%$ & $86.29\%$ & $82.67\%$ \\\hline
  obj1   & $-1.82\%$ & $-2.57\%$ & $\mathbf{-2.90\%}$ & $\mathbf{-2.90\%}$ & $-1.87\%$ & $30.92\%$ & $94.47\%$ & $90.31\%$ & $88.51\%$ \\\hline
  obj2   & $0.45\%$ & $-2.81\%$ & $\mathbf{-3.35\%}$ & $-2.49\%$ & $0.70\%$ & $57.52\%$ & $110.70\%$ & $98.68\%$ & $91.39\%$ \\\hline
  paper1 & $-3.92\%$ & $\mathbf{-7.87\%}$ & $-7.13\%$ & $-6.76\%$ & $-3.75\%$ & $52.56\%$ & $92.73\%$ & $83.92\%$ & $80.90\%$ \\\hline
  paper2 & $-3.73\%$ & $\mathbf{-8.30\%}$ & $-7.51\%$ & $-7.29\%$ & $-3.71\%$ & $52.51\%$ & $87.65\%$ & $80.64\%$ & $78.01\%$ \\\hline
  paper3 & $-3.79\%$ & $\mathbf{-7.71\%}$ & $-7.12\%$ & $-6.88\%$ & $-3.58\%$ & $45.50\%$ & $82.86\%$ & $76.56\%$ & $74.97\%$ \\\hline
  paper4 & $-4.73\%$ & $-6.83\%$ & $\mathbf{-6.87\%}$ & $-5.80\%$ & $-4.73\%$ & $39.33\%$ & $82.93\%$ & $76.98\%$ & $76.67\%$ \\\hline
  paper5 & $-2.47\%$ & $\mathbf{-5.02\%}$ & $-4.82\%$ & $-4.28\%$ & $-2.64\%$ & $40.44\%$ & $87.77\%$ & $81.47\%$ & $80.31\%$ \\\hline
  paper6 & $-3.34\%$ & $\mathbf{-6.69\%}$ & $-6.54\%$ & $-5.51\%$ & $-3.14\%$ & $52.18\%$ & $94.99\%$ & $86.38\%$ & $83.73\%$ \\\hline
  pic    & $-7.00\%$ & $\mathbf{-10.28\%}$ & $-9.26\%$ & $-8.99\%$ & $-6.95\%$ & $33.98\%$ & $76.28\%$ & $70.55\%$ & $69.26\%$ \\\hline
  progc  & $-5.15\%$ & $\mathbf{-8.17\%}$ & $-6.45\%$ & $-6.57\%$ & $-4.89\%$ & $56.14\%$ & $97.74\%$ & $88.21\%$ & $84.10\%$ \\\hline
  progl  & $-3.13\%$ & $\mathbf{-8.45\%}$ & $-8.28\%$ & $-6.84\%$ & $-3.35\%$ & $77.42\%$ & $115.20\%$ & $102.28\%$ & $94.28\%$ \\\hline
  progp  & $-4.30\%$ & $\mathbf{-7.39\%}$ & $-6.98\%$ & $-5.62\%$ & $-4.49\%$ & $81.27\%$ & $122.00\%$ & $107.69\%$ & $101.63\%$ \\\hline
  trans  & $-4.86\%$ & $-8.26\%$ & $\mathbf{-8.33\%}$ & $-6.09\%$ & $-4.17\%$ & $96.93\%$ & $142.82\%$ & $125.76\%$ & $112.64\%$ \\\hline
		\multicolumn{10}{|c|}{Canterbury corpus}
        \\\hline
  alice29.txt  & $-3.27\%$ & $\mathbf{-7.62\%}$ & $-6.63\%$ & $-6.43\%$ & $-3.48\%$ & $56.73\%$ & $88.97\%$ & $82.33\%$ & $79.56\%$ \\\hline
  asyoulik.txt & $-3.88\%$ & $\mathbf{-7.71\%}$ & $-6.03\%$ & $-5.76\%$ & $-3.56\%$ & $50.22\%$ & $84.16\%$ & $78.47\%$ & $76.91\%$ \\\hline
  cp.html      & $-5.26\%$ & $\mathbf{-7.35\%}$ & $-6.98\%$ & $-6.22\%$ & $-4.58\%$ & $50.36\%$ & $97.67\%$ & $89.16\%$ & $83.60\%$ \\\hline
  fields.c     & $-2.00\%$ & $-4.43\%$ & $-4.31\%$ & $\mathbf{-4.62\%}$ & $-2.13\%$ & $69.20\%$ & $115.19\%$ & $102.61\%$ & $97.14\%$ \\\hline
  grammar.lsp  & $-4.66\%$ & $\mathbf{-5.93\%}$ & $-4.24\%$ & $-4.80\%$ & $-3.25\%$ & $51.27\%$ & $99.01\%$ & $90.25\%$ & $86.44\%$ \\\hline
  kennedy.xls  & $-1.24\%$ & $\mathbf{-1.36\%}$ & $-0.86\%$ & $-1.32\%$ & $-1.27\%$ & $10.39\%$ & $89.53\%$ & $88.52\%$ & $88.34\%$ \\\hline
  lcet10.txt   & $-2.01\%$ & $\mathbf{-7.68\%}$ & $-6.70\%$ & $-5.65\%$ & $-2.35\%$ & $67.99\%$ & $96.57\%$ & $88.35\%$ & $83.84\%$ \\\hline
  plrabn12.txt & $-2.91\%$ & $\mathbf{-7.45\%}$ & $-6.16\%$ & $-5.20\%$ & $-2.70\%$ & $52.27\%$ & $81.47\%$ & $76.66\%$ & $74.72\%$ \\\hline
  sum          & $-1.72\%$ & $\mathbf{-3.75\%}$ & $-3.25\%$ & $-3.28\%$ & $-1.52\%$ & $44.56\%$ & $105.36\%$ & $96.07\%$ & $92.98\%$ \\\hline
  xargs.1      & $-4.85\%$ & $-6.22\%$ & $-6.54\%$ & $\mathbf{-6.65\%}$ & $-4.85\%$ & $41.77\%$ & $89.03\%$ & $82.07\%$ & $80.49\%$ \\\hline
  \end{tabular}
  \caption{Factor counts relative to LZD\@. 
	The smallest factor counts have been marked bold.
}
\label{tab:calgary}
\end{table}

It seems not straightforward to adapt the flexible parsing variant for LZW introduced by Horspool~\cite{horspool95effect} to other flexible parsing schemes like LZDR because the longest factor that can be parsed is added to the dictionary, while we only advance in the input text by the length of the factor that was chosen in a semi-greedy way.
Here, we propose two variations: 
\begin{itemize}
    \item \AltFlex{}-LZDR maintains those factors in its dictionary that are actually in the final parsing.
    \item \MaxAlt{}-LZDR adds the longest greedy factor to its dictionary, while advancing by the length of the factor selected in a semi-greedy way.
\end{itemize}

\block{Alternative Flexible LZDR}
The \AltFlex{}-LZDR parsing of $T$ is $T = F_1 F_2 \cdots F_z$ such that, for every $x \in [1..z]$, $F_x$ is chosen as the non-greedy factor using the production rules of LZDR that maximizes the combined length $|F_x G_{x+1}|$ of $F_x$ and the lookahead factor $G_{x+1}$.
Here, $G_{x+1}$ is again the lookahead factor that is chosen greedily as the next longest factor.
The factor $F_x$ can only reference the factors $F_0, F_1, F_2, \ldots, F_{x-1}$. 
That means, it can only reference previous factors that were parsed by \AltFlex{}-LZDR\@.
In the same sense, the lookahead factor $G_{x+1}$ can reference the factor $F_x$ in addition.

\block{\MaxAlt{}-LZDR}
The \MaxAlt{}-LZDR parsing of $T$ is $T = F_1 F_2 \cdots F_z$ such that, for every $x \in [1..z]$, $F_x$ is chosen as the non-greedy factor using the production rules of LZDR that maximizes the combined length $|F_x G_{x+1}|$ of $F_x$ and the lookahead factor $G_{x+1}$.
Here, $G_{x+1}$ is again the lookahead factor that is chosen greedily as the next longest factor.
The longest factors that are in the dictionary at the point of parsing $F_x$ are $\{ F_0, R_1, R_2, \ldots, R_{x-1} \}$, where for every $y \in [1..x-1]$, the starting position of $R_y$ is $\ibeg(R_y) = |F_1 \cdots F_{y-1}| + 1$ and 
$R_y$ is defined to be the longest LZDR factor greedily parsed as a prefix of $T[\ibeg(R_y)..]$.
The factor $F_x$ can only reference factors $R_y \in \{ F_0, R_1, R_2, \ldots, R_{x-1} \}$ with $\iend(R_y) < \ibeg(F_x)$.
That means, $F_x$ can only reference a factor $R_y$ if that factor $R_y$ ends before the factor $F_x$ starts.
In the same sense, the lookahead factor $G_{x+1}$ can reference the factor $R_x$ in addition, if $\iend(R_x) < \ibeg(F_x) + |F_x|$, where $R_x$ is defined as the greedily parsed factor with starting position $\ibeg(F_x)$.
Similarly, for each $w \in [1..x]$, the longest factor $R_w$ can only reference factors $R_y \in \{ F_0, R_1, R_2, \ldots, R_{w-1} \}$ with $\iend(R_y) < |F_1 \cdots F_{w-1}| + 1$.
See \cref{tab:flexEx} for an example.

Intuitively, \MaxAlt{}-LZDR transfers the idea of Horspool~\cite{horspool95effect} to LZDR: it advances by the semi-greedy factor~$F_x$, but inserts into its dictionary the longest (greedy) factor~$R_x$ starting at the same position, which becomes usable as soon as the parsing has passed the end of~$R_x$.
\AltFlex{}-LZDR references only its own previous factors, like the greedy parsing.
\StdFlex{}-LZDR and \MaxAlt{}-LZDR, in contrast, reference greedy factors~$R_y$ that need not occur in their output, so a decoder has to recompute these factors on the already decoded text, which costs additional time and space.
Only \StdFlex{}-LZDR has the guarantee of \cref{prop:flex}.
Our implementations of all three variants evaluate the lookahead alternatives explicitly; we do not know a linear-time algorithm for any of them.

\begin{table}[t]
  \caption{Comparison between LZDR and flexible parsings of the string \texttt{aaababaaaaaabaaab}}
	\label{tab:flexEx}
  \tiny
  \begin{subtable}{\linewidth}
    \centering
      \caption{LZDR (i.e., the greedy standard variant of LZDR)}
      \begin{tabular}{|c|c|c|c|c|c|}
        \hline
        $\boldsymbol{F_1}$ & $\boldsymbol{F_2}$ & $\boldsymbol{F_3}$ & $\boldsymbol{F_4}$ & $\boldsymbol{F_5}$ & $\boldsymbol{F_6}$ \\\hline
        \texttt{aaa} & \texttt{ba} & \texttt{baaaa} & \texttt{aa} & \texttt{baaa} & \texttt{b} \\\hline
        $\texttt{a}^3$ & $(\texttt{b}, \texttt{a})$ & $(F_2, F_1)$ & $(\texttt{a}, \texttt{a})$ & $(F_2, F_4)$ & $(\texttt{b}, F_0)$ \\\hline
      \end{tabular}
  \end{subtable}

  \begin{subtable}{\linewidth}
    \centering
		\caption{Standard Flexible LZDR (\StdFlex{}-LZDR)}
      \begin{tabular}{|c|c|c|c|c|}
        \hline
        $\boldsymbol{F_1}$ & $\boldsymbol{F_2}$ & $\boldsymbol{F_3}$ & $\boldsymbol{F_4}$ & $\boldsymbol{F_5}$ \\\hline
        \texttt{aaa} & \texttt{ba} & \texttt{baaa} & \texttt{aaaba} & \texttt{aab} \\\hline
        $\texttt{a}^3$ & $(\texttt{b}, \texttt{a})$ & $(R_2, R_1)[1..4]$ & $(R_1, R_2)$ & $(R_4, \texttt{b})$ \\\hline
      \end{tabular}
  \end{subtable}

  \begin{subtable}{\linewidth}
    \centering
		\caption{\AltFlex{}-LZDR}
      \begin{tabular}{|c|c|c|c|c|}
        \hline
        $\boldsymbol{F_1}$ & $\boldsymbol{F_2}$ & $\boldsymbol{F_3}$ & $\boldsymbol{F_4}$ & $\boldsymbol{F_5}$ \\\hline
        \texttt{aaa} & \texttt{ba} & \texttt{baaa} & \texttt{aaabaaa} & \texttt{b} \\\hline
        $\texttt{a}^3$ & $(\texttt{b}, \texttt{a})$ & $(F_2, F_1)[1..4]$ & $(F_1, F_3)$ & $(\texttt{b}, F_0)$ \\\hline
      \end{tabular}
  \end{subtable}

  \begin{subtable}{\linewidth}
    \centering
		\caption{\MaxAlt{}-LZDR}
      \begin{tabular}{|c|c|c|c|c|}
        \hline
        $\boldsymbol{F_1}$ & $\boldsymbol{F_2}$ & $\boldsymbol{F_3}$ & $\boldsymbol{F_4}$ & $\boldsymbol{F_5}$ \\\hline
        \texttt{aaa} & \texttt{ba} & \texttt{baaa} & \texttt{aaab} & \texttt{aaab} \\\hline
        $\texttt{a}^3$ & $(\texttt{b}, \texttt{a})$ & $(R_2, R_1)[1..4]$ & $(R_1, R_2)[1..4]$ & $(R_1, \texttt{b})$ \\\hline
        $\boldsymbol{R_1}$ & $\boldsymbol{R_2}$ & $\boldsymbol{R_3}$ & $\boldsymbol{R_4}$ & $\boldsymbol{R_5}$ \\\hline
        \texttt{aaa} & \texttt{ba} & \texttt{baaaa} & \texttt{aaaba} & \texttt{aaab} \\\hline
        $\texttt{a}^3$ & $(\texttt{b}, \texttt{a})$ & $(R_2, R_1)$ & $(R_1, R_2)$ & $(R_1, \texttt{b})$ \\\hline
      \end{tabular}
  \end{subtable}
\end{table}

\section{Bounds on the number of factors}\label{sec:bounds}

In this section, we compare the numbers of factors of the greedy LZD, LZD+, and LZDR factorizations of~$T$, which we denote by $z_{\mathrm{LZD}}$, $z_+$, and $z_R$, with the size~$g$ of a smallest SLP generating~$T$ (\cref{sec:preliminaries}); we write $z_R(T)$ etc.\ when the text matters.
Counting factors suffices up to constant factors: an LZD factorization induces an SLP with $O(z_{\mathrm{LZD}})$ rules, and an LZD+ or LZDR factorization induces a collage system with $O(z_+)$ or $O(z_R)$ assignments (\cref{sec:collage}).
Our main result is a family of binary strings, cf.~\cref{fig:binary}, on which all three schemes produce $\Omega(n^{2/3})$ factors although a grammar of size $O(n^{1/3})$ exists (\cref{sec:bounds-lower}); hence, neither truncations nor repetitions improve the approximation ratio of LZD.
\Cref{sec:bounds-upper} complements this result with upper bounds.

\subsection{A binary lower bound for all three schemes}\label{sec:bounds-lower}

On the strings~$S_k$ of Badkobeh et al.\ (\cref{sec:lowerbound}), LZDR produces far fewer factors than LZD and LZD+ (\cref{tab:lowerBound}).
This advantage depends on the position at which the parsing enters a copy of the hard part, and it disappears for a suitable rotation of the hard part.
Combining such a rotation with a binary training prefix, we obtain strings on which all three schemes produce the same factors, arranged in rows as described for LZD in \cref{sec:lowerbound}.

Throughout this section, $k \ge 16$ is a power of two, $m := k/8$, and $\tau := k/2+1$.
We use the strings $\Delta_i := \texttt{a}^i\texttt{bb}\texttt{a}^{k-i}$ of length $k+2$ from \cref{sec:lowerbound}, which differ only in the position of~$\texttt{bb}$, and their prefixes $\Gamma_i := \texttt{a}^i\texttt{bb}$ for $0 \le i \le k$.
The binary \emph{training prefix}
\[
  U_k := \texttt{bb}\,\texttt{aa}
  \Bigl(\prod_{r=3}^{k}\texttt{b}^r\texttt{a}^r\Bigr)
  \texttt{b}^{k+1}
  \Bigl(\prod_{i=\tau}^{k}\Gamma_i\Bigr)
  \Bigl(\prod_{i=\tau}^{k-1}\Delta_i\,\texttt{b}^{k/2+i+1}\Bigr),
\]
where the products denote concatenations in increasing order of the index, first creates runs of $\texttt{a}$ and~$\texttt{b}$ up to length~$k$, then the strings~$\Gamma_i$, and then the strings~$\Delta_i$ for $i \ge \tau$, each followed by a separating run of~$\texttt{b}$ that is longer than all earlier runs of~$\texttt{b}$.
The \emph{hard part}
\[
  V_{k,m} := \Delta_k\Delta_{k-1}\,\Delta_k\Delta_{k-2}\cdots\Delta_k\Delta_{k-m+1}\,\Delta_k\texttt{a}^{k-1}
\]
lists the pairs $\Delta_k\Delta_{k-j}$ for $j = 1, \ldots, m-1$, followed by $\Delta_k\texttt{a}^{k-1}$, so $|V_{k,m}| = (2k+4)m-3$; for $m = k/2$, 
$V_{k,k/2}$ is the hard part~$V_k$ of~$S_k$ defined in \cref{sec:lowerbound}.
For a string~$S$ and $s \in [0..|S|]$, let $\operatorname{rot}_s(S) := S[s+1..|S|]\,S[1..s]$ denote the cyclic shift of $S$ by $s$ letters to the left, 
and let $V'_{k,m} := \operatorname{rot}_m(V_{k,m})$ be the hard part shifted by $m$ letters.
We study the binary strings
\(
  B_k := U_k\,(V'_{k,m})^m,
\)
whose \emph{hard suffix} $(V'_{k,m})^m$ consists of $m$ copies of the shifted hard part.

Besides its alphabet, $B_k$ differs from $S_k = P_k V_k^{k/2}$ in three respects.
First, its training prefix separates blocks by runs of~$\texttt{b}$ instead of $\texttt{c}$ and~$\texttt{d}$, and it creates only the strings~$\Delta_i$ with $i \ge \tau$.
Second, its hard part has only $m-1 = k/8-1$ pairs instead of $k/2-1$, and its hard suffix has only $m$ instead of $k/2$ copies, so that the factors of the hard suffix use only these strings~$\Delta_i$.
Third, each copy of the hard part is rotated by $m$ letters, so that the parsing enters each copy at least $m$ letters behind its beginning, instead of $0, 1, 2, \ldots$ letters behind it as on~$S_k$; this keeps LZDR from escaping the regular pattern of factors that LZD follows (\cref{fig:candidates}).
As a result, the hard suffix still costs $\Theta(k^2)$ factors, namely $k^2/64$, but now for all three schemes.

Our first observation explains why LZD and LZD+ have the same number of factors on~$S_k$ (\cref{tab:lowerBound}) and on~$B_k$: LZD+ can deviate from LZD only by truncating a reference.

\begin{lemma}\label[lemma]{lem:lzd-coincide}
Suppose that every factor of the LZD+ factorization of~$T$ can be written as $F_x = F_{x_1}F_{x_2}$, where $F_{x_1}$ is the longest complete reference as in the combination rule and $F_{x_2} \in \{F_0, F_1, \ldots, F_{x-1}\} \cup \Sigma$ is not truncated.
Then the LZD factorization of~$T$ is the same.
\end{lemma}
\begin{proof}

By induction on~$x$, both schemes have the same dictionary when they compute~$F_x$, and hence the same first component~$F_{x_1}$.
Let $G$ be the longest string in $\{F_0, F_1, \ldots, F_{x-1}\} \cup \Sigma$ that is a prefix of the text after~$F_{x_1}$, so that LZD outputs $F_{x_1}G$.
Since $F_{x_2}$ is such a string, $|G| \ge |F_{x_2}|$; since $F_{x_1}G$ is an LZD+ candidate, $|F_{x_1}G| \le |F_x|$.
Hence, $F_{x_1}G = F_x$.

\end{proof}

For the smallest instance $k = 16$, we have $m = 2$ and $\tau = 9$, so $V_{16,2} = \Delta_{16}\Delta_{15}\,\Delta_{16}\texttt{a}^{15}$ of length $|V_{16,2}| = 69$, and
\[
  B_{16} = U_{16}\,(V'_{16,2})^2
  \quad\text{with}\quad
  V'_{16,2} = \operatorname{rot}_2(V_{16,2}) = \texttt{a}^{14}\texttt{bb}\,\texttt{a}^{15}\texttt{bb}\,\texttt{a}^{17}\texttt{bb}\,\texttt{a}^{17},
\]
where $|U_{16}| = 676$ and $|B_{16}| = 814$.
All three schemes factorize $B_{16}$ into $7k/2-3 = 53$ factors of the training prefix and the $m^2 = 4$ factors
$ \Delta_{14}\Delta_{13}, \Delta_{14}\texttt{a}^{16}, \Delta_{13}\Delta_{12},$ and $\Delta_{13}\texttt{a}^{14} $
of the hard suffix, which are two rows of two factors: the first row starts at the beginning of the hard suffix and the second row one letter further inside the second copy of~$V'_{16,2}$, so that the strings~$\Delta_i$ it reads have indices one smaller (\cref{fig:rows} shows the same pattern for $k = 32$).

\begin{theorem}\label[theorem]{thm:binary}

For every power of two $k \ge 16$, LZD, LZD+, and LZDR produce the same factorization of~$B_k$, with $z_{\mathrm{LZD}}(B_k) = z_+(B_k) = z_R(B_k) = 7k/2-3+k^2/64$ factors, whereas $|B_k| = \Theta(k^3)$ and $g(B_k) = O(k)$.
Hence, the approximation ratio of each of the three schemes is $\Omega(n^{1/3})$, already for binary strings.

\end{theorem}

We prove the theorem in two steps, which we treat in turn: all three schemes parse the training prefix~$U_k$ block by block (\cref{lem:binary-prefix}), and they parse the hard suffix into $m$ \emph{rows} of $m$ factors each (\cref{lem:rows}), as in the example above.
Throughout, we use that $\Gamma_k = \Delta_k$ and that $\Delta_i = \Gamma_i\texttt{a}^{k-i}$.

\begin{lemma}\label[lemma]{lem:binary-prefix}

For each of LZD, LZD+, and LZDR, the factorization of~$U_k$ is
\[
  \texttt{bb},\texttt{aa},\texttt{b}^3,\texttt{a}^3,\ldots,\texttt{b}^k,\texttt{a}^k,\texttt{b}^{k+1},
  \qquad
  \Gamma_\tau,\Gamma_{\tau+1},\ldots,\Gamma_k,
  \qquad
  \Delta_\tau,\texttt{b}^{k+2},\Delta_{\tau+1},\texttt{b}^{k+3},\ldots,\Delta_{k-1},\texttt{b}^{3k/2},
\]
with $(2k-1) + k/2 + (k-2) = 7k/2-3$ factors.
Afterwards, the runs~$\texttt{a}^i$ ($2 \le i \le k$) and the strings $\Gamma_i$ and~$\Delta_i$ ($\tau \le i \le k$) are factors, and every factor longer than $k+2$ is a run of~$\texttt{b}$.
Moreover, $|U_k| = O(k^2)$ and $g(U_k) = O(k)$.

\end{lemma}
\begin{proof}

The first part consists of runs whose lengths increase by one, starting with $\texttt{bb} = \texttt{b}\cdot\texttt{b}$ and $\texttt{aa} = \texttt{a}\cdot\texttt{a}$; each run is the previous run of the same letter extended by this letter, and no candidate extends beyond the end of a run, since all previous factors are runs.

For $\tau \le i \le k$, the longest complete reference at the start of~$\Gamma_i$ is~$\texttt{a}^i$, since each earlier~$\Gamma_j$ has its $\texttt{b}$ too early.
After~$\texttt{a}^i$, the text starts with $\texttt{bba}$, whereas all previous factors starting with~$\texttt{b}$ are runs of~$\texttt{b}$; so the second component is~$\texttt{bb}$.
Truncations and repetitions of earlier factors break off at the first~$\texttt{b}$ or earlier.
Hence, the next factor is~$\Gamma_i$.

For $\tau \le i < k$, the block $\Delta_i = \Gamma_i\texttt{a}^{k-i}$ starts with the longest complete reference~$\Gamma_i$.
Since $i > k/2$, we have $k-i < \tau$, so after~$\Gamma_i$, the text has $k-i$ letters~$\texttt{a}$ followed by~$\texttt{b}$, whereas every previous factor starting with~$\texttt{a}$ is a run of~$\texttt{a}$ or has its first~$\texttt{b}$ only after at least~$\tau$ letters~$\texttt{a}$.
Hence, the second component is~$\texttt{a}^{k-i}$ (the letter~$\texttt{a}$ for $i = k-1$), and the combination is~$\Delta_i$.
For LZDR, the repetition of~$\Gamma_i$ matches $\texttt{a}^{k-i}$ after its first copy and then breaks off at the separator, which yields~$\Delta_i$ again; all other repetitions are shorter.

Each separator $\texttt{b}^{k/2+i+1}$ is one letter longer than the previous run of~$\texttt{b}$, so it is the combination of this run and the letter~$\texttt{b}$.
No candidate extends into the following~$\texttt{a}$, since every previous factor starting with~$\texttt{b}$ is a run of~$\texttt{b}$.

Every factor is the concatenation of the longest complete reference and a complete reference, so by \cref{lem:lzd-coincide}, the factorization is the same for LZD and LZD+.
All runs in~$U_k$ can be generated with $O(k)$ rules, after which each~$\Gamma_i$ and each~$\Delta_i$ needs $O(1)$ further rules; since $U_k$ consists of $O(k)$ blocks of length $O(k)$ each, $|U_k| = O(k^2)$ and $g(U_k) = O(k)$.

\end{proof}

\begin{figure}[t]

\centering
\begin{tikzpicture}[y=1cm,font=\scriptsize]
\node[anchor=east] at (-0.15,{0.22}) {$S_k = $};
\draw[fill=white] (0,{0}) rectangle (0.57,{0.44});
\node at ({0.285},{0.22}) {$\texttt{a}^2$};
\draw[fill=black!18] (0.57,{0}) rectangle (1.14,{0.44});
\node at ({0.855},{0.22}) {$\texttt{c}^2$};
\draw[fill=white] (1.14,{0}) rectangle (1.71,{0.44});
\node at ({1.425},{0.22}) {$\texttt{a}^3$};
\draw[fill=black!18] (1.71,{0}) rectangle (2.28,{0.44});
\node at ({1.995},{0.22}) {$\texttt{c}^3$};
\node at ({2.53},{0.22}) {$\cdots$};
\draw[fill=white] (2.78,{0}) rectangle (3.35,{0.44});
\node at ({3.065},{0.22}) {$\texttt{a}^k$};
\draw[fill=black!18] (3.35,{0}) rectangle (3.92,{0.44});
\node at ({3.635},{0.22}) {$\texttt{c}^k$};
\draw[fill=white] (4.17,{0}) rectangle (4.74,{0.44});
\node at ({4.455},{0.22}) {$\Gamma_0$};
\draw[fill=white] (4.74,{0}) rectangle (5.31,{0.44});
\node at ({5.025},{0.22}) {$\Gamma_1$};
\node at ({5.56},{0.22}) {$\cdots$};
\draw[fill=white] (5.81,{0}) rectangle (6.57,{0.44});
\node at ({6.19},{0.22}) {$\Gamma_{k-1}$};
\draw[fill=white] (6.82,{0}) rectangle (7.39,{0.44});
\node at ({7.105},{0.22}) {$\Delta_0$};
\draw[fill=black!18] (7.39,{0}) rectangle (7.96,{0.44});
\node at ({7.675},{0.22}) {$\texttt{d}^2$};
\draw[fill=white] (7.96,{0}) rectangle (8.53,{0.44});
\node at ({8.245},{0.22}) {$\Delta_1$};
\draw[fill=black!18] (8.53,{0}) rectangle (9.1,{0.44});
\node at ({8.815},{0.22}) {$\texttt{d}^3$};
\node at ({9.35},{0.22}) {$\cdots$};
\draw[fill=white] (9.6,{0}) rectangle (10.17,{0.44});
\node at ({9.885},{0.22}) {$\Delta_k$};
\draw[fill=black!18] (10.17,{0}) rectangle (10.93,{0.44});
\node at ({10.55},{0.22}) {$\texttt{d}^{k+2}$};
\draw[fill=orange!25] (11.18,{0}) rectangle (12.58,{0.44});
\node at ({11.88},{0.22}) {$V_k^{k/2}$};
\draw (0.03,{0.5}) -- ++(0,0.08) -- (3.89,{0.58}) -- ++(0,-0.08);
\node[above] at ({1.96},{0.55}) {first part};
\draw (4.2,{0.5}) -- ++(0,0.08) -- (6.54,{0.58}) -- ++(0,-0.08);
\node[above] at ({5.37},{0.55}) {second part};
\draw (6.85,{0.5}) -- ++(0,0.08) -- (10.9,{0.58}) -- ++(0,-0.08);
\node[above] at ({8.875},{0.55}) {third part};
\draw (11.21,{0.5}) -- ++(0,0.08) -- (12.55,{0.58}) -- ++(0,-0.08);
\node[above] at ({11.88},{0.55}) {hard suffix};
\node[anchor=east] at (-0.15,{-1.13}) {$B_k = $};
\draw[fill=white] (0,{-1.35}) rectangle (0.57,{-0.91});
\node at ({0.285},{-1.13}) {$\texttt{bb}$};
\draw[fill=white] (0.57,{-1.35}) rectangle (1.14,{-0.91});
\node at ({0.855},{-1.13}) {$\texttt{aa}$};
\draw[fill=black!18] (1.14,{-1.35}) rectangle (1.71,{-0.91});
\node at ({1.425},{-1.13}) {$\texttt{b}^3$};
\draw[fill=white] (1.71,{-1.35}) rectangle (2.28,{-0.91});
\node at ({1.995},{-1.13}) {$\texttt{a}^3$};
\node at ({2.53},{-1.13}) {$\cdots$};
\draw[fill=black!18] (2.78,{-1.35}) rectangle (3.35,{-0.91});
\node at ({3.065},{-1.13}) {$\texttt{b}^k$};
\draw[fill=white] (3.35,{-1.35}) rectangle (3.92,{-0.91});
\node at ({3.635},{-1.13}) {$\texttt{a}^k$};
\draw[fill=black!18] (3.92,{-1.35}) rectangle (4.68,{-0.91});
\node at ({4.3},{-1.13}) {$\texttt{b}^{k+1}$};
\draw[fill=white] (4.93,{-1.35}) rectangle (5.5,{-0.91});
\node at ({5.215},{-1.13}) {$\Gamma_\tau$};
\node at ({5.75},{-1.13}) {$\cdots$};
\draw[fill=white] (6,{-1.35}) rectangle (6.57,{-0.91});
\node at ({6.285},{-1.13}) {$\Gamma_k$};
\draw[fill=white] (6.82,{-1.35}) rectangle (7.39,{-0.91});
\node at ({7.105},{-1.13}) {$\Delta_\tau$};
\draw[fill=black!18] (7.39,{-1.35}) rectangle (8.15,{-0.91});
\node at ({7.77},{-1.13}) {$\texttt{b}^{k+2}$};
\node at ({8.4},{-1.13}) {$\cdots$};
\draw[fill=white] (8.65,{-1.35}) rectangle (9.41,{-0.91});
\node at ({9.03},{-1.13}) {$\Delta_{k-1}$};
\draw[fill=black!18] (9.41,{-1.35}) rectangle (10.17,{-0.91});
\node at ({9.79},{-1.13}) {$\texttt{b}^{3k/2}$};
\draw[fill=orange!25] (10.42,{-1.35}) rectangle (11.82,{-0.91});
\node at ({11.12},{-1.13}) {$(V'_{k,m})^m$};
\draw (0.03,{-0.85}) -- ++(0,0.08) -- (4.65,{-0.77}) -- ++(0,-0.08);
\node[above] at ({2.34},{-0.8}) {first part};
\draw (4.96,{-0.85}) -- ++(0,0.08) -- (6.54,{-0.77}) -- ++(0,-0.08);
\node[above] at ({5.75},{-0.8}) {second part};
\draw (6.85,{-0.85}) -- ++(0,0.08) -- (10.14,{-0.77}) -- ++(0,-0.08);
\node[above] at ({8.495},{-0.8}) {third part};
\draw (10.45,{-0.85}) -- ++(0,0.08) -- (11.79,{-0.77}) -- ++(0,-0.08);
\node[above] at ({11.12},{-0.8}) {hard suffix};
\end{tikzpicture}
\caption{Setting of \cref{lem:binary-prefix}: the training prefixes of~$S_k$ (top) and of the binary string~$B_k$ (bottom), not to scale. 
Each box is one factor of LZD, LZD+, and LZDR; gray boxes are separators, which are never referenced in the hard suffix. 
In~$B_k$, runs of~$\texttt{b}$ take the role of the separators $\texttt{c}^i$ and~$\texttt{d}^i$, and only the strings~$\Delta_i$ with $i \ge \tau = k/2+1$ are created; they suffice for the factors of the hard suffix (\cref{fig:rows}). 
Separators longer than $k+2$ become long factors, but they start with~$\texttt{b}$, whereas every factor of the hard suffix and its second component start with~$\texttt{a}$.}
\label{fig:binary}

\end{figure}

\block{Phases}
Before we turn to the hard suffix, we fix how we address its positions.
Let $W := V_{k,m}^\infty$.
The hard suffix of~$B_k$ is $W[m+1..m+m|V_{k,m}|]$, i.e., it starts $m$ letters behind the beginning of a copy of~$V_{k,m}$ in~$W$.
A position of the hard suffix has \emph{phase}~$r$ if it lies $r$ letters behind the beginning of a copy of~$V_{k,m}$ in~$W$; thus, the hard suffix starts at phase~$m$, and phases are taken modulo~$|V_{k,m}|$.

\begin{lemma}\label[lemma]{lem:rows}

For each of LZD, LZD+, and LZDR, the hard suffix of~$B_k$ is parsed into $m$ rows: for $r = m, m+1, \ldots, 2m-1$ and $p := k-r$, the row starting at phase~$r$ consists of the factors $A_{r,j} := \Delta_p\Delta_{p-j}$ for $1 \le j < m$, followed by $A_{r,m} := \Delta_p\texttt{a}^k$ if $r < 2m-1$, and by the remaining suffix $\Delta_p\texttt{a}^{k-m}$ if $r = 2m-1$.

\end{lemma}
\begin{proof}

Shifting the text by $r$ letters turns each block~$\Delta_i$ into~$\Delta_{i-r}$, since it moves $r$ letters~$\texttt{a}$ from the beginning of the block to its end; this is why a row starting at phase~$r$ reads the pairs $\Delta_k\Delta_{k-j}$ of~$V_{k,m}$ as $\Delta_p\Delta_{p-j}$ (\cref{fig:rows}).
We proceed by induction on~$r$ and consider LZD+ and LZDR; LZD is covered at the end.
Suppose that the parsing has reached phase~$r$ of the hard suffix, and that the previous factors are those of~$U_k$ and those of the rows starting at the phases $m, \ldots, r-1$.
Since $m \le r \le 2m-1$, every index $p-j$ with $j < m$ satisfies $p-j \ge k-3m+2 > \tau$, so $\Delta_p$, $\Delta_{p-j}$, and~$\Gamma_p$ are factors of~$U_k$ by \cref{lem:binary-prefix}.

For $j < m$, the $j$th factor of the row starts at phase $r + (2k+4)(j-1)$, that is, $r$ letters inside the occurrence of~$\Delta_k$ that precedes~$\Delta_{k-j}$ in~$V_{k,m}$.
Since $r < 2m < k-j$, the next $2k+4$ letters are
\[
  \texttt{a}^{k-r}\texttt{bb}\texttt{a}^r\cdot\texttt{a}^{k-r-j}\texttt{bb}\texttt{a}^{r+j} = \Delta_p\Delta_{p-j},
\]
where the last $r$ letters belong to the next occurrence of~$\Delta_k$.

The longest complete reference at this position is~$\Delta_p$.
Indeed, factors of~$U_k$ that start with~$\texttt{a}$ have length at most $k+2$, and among them only~$\Delta_p$ matches $k+2$ letters; every factor of an earlier row starts with some~$\Delta_{p'}$ with $p' > p$, whose first~$\texttt{b}$ follows after more than $p$ letters~$\texttt{a}$, so it mismatches within the first $k+2$ letters; and every earlier factor~$A_{r,q}$ of the current row, $q < j$, continues after~$\Delta_p$ with~$\Delta_{p-q}$ instead of~$\Delta_{p-j}$, whose $\texttt{bb}$ comes $j-q$ positions later, so it mismatches within the first $2k+4$ letters.
After~$\Delta_p$, a previous factor that matches more than $k+2$ letters would have to start with~$\Delta_{p-j}$; but factors of~$U_k$ starting with~$\texttt{a}$ have at most $k+2$ letters, and factors of the rows start with some~$\Delta_{p'}$ with $p' \ge p$.
Hence, the longest combination is~$A_{r,j}$, of length $2k+4$, and no truncation is longer, since every previous factor longer than $k+2$ mismatches within the first $2k+4$ letters.

For LZDR, consider the repetition of a previous factor or letter~$C$.
If the first copy of~$C$ does not match completely, the repetition is shorter than $2k+4$ by the above.
Otherwise, $C$ is a letter, a run of~$\texttt{a}$, $\Gamma_p$, or~$\Delta_p$, since the text starts with $\texttt{a}^p\texttt{bb}$.
The repetition of a letter or of a run of~$\texttt{a}$ breaks off at the first~$\texttt{b}$.
After its first copy, the repetition of~$\Delta_p$ expects $\texttt{a}^p\texttt{bb}$ but meets the~$\texttt{bb}$ of~$\Delta_{p-j}$ after $p-j$ letters~$\texttt{a}$, so it covers $k+2+p-j < 2k+4$ letters.
After its first copy $\texttt{a}^p\texttt{bb}$, the repetition of~$\Gamma_p$ meets $k-j$ letters~$\texttt{a}$ before the next~$\texttt{b}$; since $j < m \le r$, we have $k-j > p$, so the second copy breaks off after $p$ letters~$\texttt{a}$, and the repetition covers $2p+2 < 2k+4$ letters.
This is the only place where we use $r \ge m$, i.e., the rotation: for $j = r$, the second copy would match completely, and the repetition would cover $3p+4 > 2k+4$ letters and thus win (\cref{fig:candidates}).
Hence, $A_{r,j}$ is also the factor of LZDR.

After $A_{r,1}, \ldots, A_{r,m-1}$, the parsing is $r$ letters inside the last occurrence of~$\Delta_k$ in the current copy of~$V_{k,m}$, which is followed by~$\texttt{a}^{k-1}$.
If $r < 2m-1$, the text continues with $\Delta_p$, the remaining $k-1-r$ letters~$\texttt{a}$ of this copy, and the block $\texttt{a}^k\texttt{bb}$ starting the next copy, i.e., with $\Delta_p\texttt{a}^{2k-1-r}\texttt{bb}$.
As above, $\Delta_p$ is the longest complete reference.
After it, the factor~$\texttt{a}^k$ matches, while the only previous factor that properly extends~$\texttt{a}^k$ is $\Delta_k = \Gamma_k$, which continues with~$\texttt{b}$; so the longest combination is $A_{r,m} = \Delta_p\texttt{a}^k$, of length $2k+2$.
An earlier factor~$A_{r,q}$ of the current row agrees after~$\Delta_p$ with only $p-q < k$ further letters, factors of earlier rows mismatch within the first $k+2$ letters, and the repetitions of~$\Delta_p$ and~$\Gamma_p$ cover only $k+2+p$ and $2p+2$ letters, both less than $2k+2$.
Hence, $A_{r,m}$ is the next factor, and the row covers $(2k+4)(m-1)+2k+2 = |V_{k,m}|+1$ letters, so the next row starts at phase~$r+1$.

If $r = 2m-1$, the text ends after~$\Delta_p$ and $k-m$ further letters~$\texttt{a}$: the hard suffix has length $m|V_{k,m}|$, the last row starts at its position $(m-1)(|V_{k,m}|+1)$, and $m|V_{k,m}| - (m-1)(|V_{k,m}|+1) - (2k+4)(m-1) - (k+2) = k-m$.
These letters form a prefix of the factor~$\texttt{a}^k$, and $\texttt{a}^{k-m}$ is itself a factor, so the combination $\Delta_p\texttt{a}^{k-m}$ ends the text.
This completes the induction for LZD+ and LZDR.
Every factor of the rows, like every factor of~$U_k$, is the concatenation of the longest complete reference and a complete reference, so, by \cref{lem:lzd-coincide}, LZD produces the same factors.

\end{proof}

\begin{figure}[t]

\centering
\begin{tikzpicture}[x=0.04201cm,y=1cm,font=\scriptsize]
\draw[black!35,densely dashed] (0,{0.37}) -- (0,{-4.3});
\draw[black!35,densely dashed] (68,{0.37}) -- (68,{-4.3});
\draw[black!35,densely dashed] (136,{0.37}) -- (136,{-4.3});
\draw[black!35,densely dashed] (204,{0.37}) -- (204,{-4.3});
\node at ({17},0.5) {$\Delta_{32}$};
\node at ({51},0.5) {$\Delta_{31}$};
\node at ({85},0.5) {$\Delta_{32}$};
\node at ({119},0.5) {$\Delta_{30}$};
\node at ({153},0.5) {$\Delta_{32}$};
\node at ({187},0.5) {$\Delta_{29}$};
\node at ({221},0.5) {$\Delta_{32}$};
\node at ({253.5},0.5) {$\texttt{a}^{31}$};
\node[anchor=east] at (-2,{0.16}) {copy 1};
\fill[black!15] (0,{0}) rectangle (4,{0.32});
\draw[black!40] (0,{0}) rectangle (269,{0.32});
\fill[black] (32,{0}) rectangle (33,{0.32});
\fill[black] (33,{0}) rectangle (34,{0.32});
\fill[black] (65,{0}) rectangle (66,{0.32});
\fill[black] (66,{0}) rectangle (67,{0.32});
\fill[black] (100,{0}) rectangle (101,{0.32});
\fill[black] (101,{0}) rectangle (102,{0.32});
\fill[black] (132,{0}) rectangle (133,{0.32});
\fill[black] (133,{0}) rectangle (134,{0.32});
\fill[black] (168,{0}) rectangle (169,{0.32});
\fill[black] (169,{0}) rectangle (170,{0.32});
\fill[black] (199,{0}) rectangle (200,{0.32});
\fill[black] (200,{0}) rectangle (201,{0.32});
\fill[black] (236,{0}) rectangle (237,{0.32});
\fill[black] (237,{0}) rectangle (238,{0.32});
\draw[black!40] (34,{0}) -- (34,{0.32});
\draw[black!40] (68,{0}) -- (68,{0.32});
\draw[black!40] (102,{0}) -- (102,{0.32});
\draw[black!40] (136,{0}) -- (136,{0.32});
\draw[black!40] (170,{0}) -- (170,{0.32});
\draw[black!40] (204,{0}) -- (204,{0.32});
\draw[black!40] (238,{0}) -- (238,{0.32});
\node[anchor=east] at (-2,{-1.09}) {copy 2};
\draw[black!40] (0,{-1.25}) rectangle (269,{-0.93});
\fill[black] (32,{-1.25}) rectangle (33,{-0.93});
\fill[black] (33,{-1.25}) rectangle (34,{-0.93});
\fill[black] (65,{-1.25}) rectangle (66,{-0.93});
\fill[black] (66,{-1.25}) rectangle (67,{-0.93});
\fill[black] (100,{-1.25}) rectangle (101,{-0.93});
\fill[black] (101,{-1.25}) rectangle (102,{-0.93});
\fill[black] (132,{-1.25}) rectangle (133,{-0.93});
\fill[black] (133,{-1.25}) rectangle (134,{-0.93});
\fill[black] (168,{-1.25}) rectangle (169,{-0.93});
\fill[black] (169,{-1.25}) rectangle (170,{-0.93});
\fill[black] (199,{-1.25}) rectangle (200,{-0.93});
\fill[black] (200,{-1.25}) rectangle (201,{-0.93});
\fill[black] (236,{-1.25}) rectangle (237,{-0.93});
\fill[black] (237,{-1.25}) rectangle (238,{-0.93});
\draw[black!40] (34,{-1.25}) -- (34,{-0.93});
\draw[black!40] (68,{-1.25}) -- (68,{-0.93});
\draw[black!40] (102,{-1.25}) -- (102,{-0.93});
\draw[black!40] (136,{-1.25}) -- (136,{-0.93});
\draw[black!40] (170,{-1.25}) -- (170,{-0.93});
\draw[black!40] (204,{-1.25}) -- (204,{-0.93});
\draw[black!40] (238,{-1.25}) -- (238,{-0.93});
\node[anchor=east] at (-2,{-2.34}) {copy 3};
\draw[black!40] (0,{-2.5}) rectangle (269,{-2.18});
\fill[black] (32,{-2.5}) rectangle (33,{-2.18});
\fill[black] (33,{-2.5}) rectangle (34,{-2.18});
\fill[black] (65,{-2.5}) rectangle (66,{-2.18});
\fill[black] (66,{-2.5}) rectangle (67,{-2.18});
\fill[black] (100,{-2.5}) rectangle (101,{-2.18});
\fill[black] (101,{-2.5}) rectangle (102,{-2.18});
\fill[black] (132,{-2.5}) rectangle (133,{-2.18});
\fill[black] (133,{-2.5}) rectangle (134,{-2.18});
\fill[black] (168,{-2.5}) rectangle (169,{-2.18});
\fill[black] (169,{-2.5}) rectangle (170,{-2.18});
\fill[black] (199,{-2.5}) rectangle (200,{-2.18});
\fill[black] (200,{-2.5}) rectangle (201,{-2.18});
\fill[black] (236,{-2.5}) rectangle (237,{-2.18});
\fill[black] (237,{-2.5}) rectangle (238,{-2.18});
\draw[black!40] (34,{-2.5}) -- (34,{-2.18});
\draw[black!40] (68,{-2.5}) -- (68,{-2.18});
\draw[black!40] (102,{-2.5}) -- (102,{-2.18});
\draw[black!40] (136,{-2.5}) -- (136,{-2.18});
\draw[black!40] (170,{-2.5}) -- (170,{-2.18});
\draw[black!40] (204,{-2.5}) -- (204,{-2.18});
\draw[black!40] (238,{-2.5}) -- (238,{-2.18});
\node[anchor=east] at (-2,{-3.59}) {copy 4};
\draw[black!40] (0,{-3.75}) rectangle (269,{-3.43});
\fill[black] (32,{-3.75}) rectangle (33,{-3.43});
\fill[black] (33,{-3.75}) rectangle (34,{-3.43});
\fill[black] (65,{-3.75}) rectangle (66,{-3.43});
\fill[black] (66,{-3.75}) rectangle (67,{-3.43});
\fill[black] (100,{-3.75}) rectangle (101,{-3.43});
\fill[black] (101,{-3.75}) rectangle (102,{-3.43});
\fill[black] (132,{-3.75}) rectangle (133,{-3.43});
\fill[black] (133,{-3.75}) rectangle (134,{-3.43});
\fill[black] (168,{-3.75}) rectangle (169,{-3.43});
\fill[black] (169,{-3.75}) rectangle (170,{-3.43});
\fill[black] (199,{-3.75}) rectangle (200,{-3.43});
\fill[black] (200,{-3.75}) rectangle (201,{-3.43});
\fill[black] (236,{-3.75}) rectangle (237,{-3.43});
\fill[black] (237,{-3.75}) rectangle (238,{-3.43});
\draw[black!40] (34,{-3.75}) -- (34,{-3.43});
\draw[black!40] (68,{-3.75}) -- (68,{-3.43});
\draw[black!40] (102,{-3.75}) -- (102,{-3.43});
\draw[black!40] (136,{-3.75}) -- (136,{-3.43});
\draw[black!40] (170,{-3.75}) -- (170,{-3.43});
\draw[black!40] (204,{-3.75}) -- (204,{-3.43});
\draw[black!40] (238,{-3.75}) -- (238,{-3.43});
\node[anchor=east] at (-2,{-4.84}) {tail};
\draw[black!40] (0,{-5}) rectangle (269,{-4.68});
\fill[black] (32,{-5}) rectangle (33,{-4.68});
\fill[black] (33,{-5}) rectangle (34,{-4.68});
\fill[black] (65,{-5}) rectangle (66,{-4.68});
\fill[black] (66,{-5}) rectangle (67,{-4.68});
\fill[black] (100,{-5}) rectangle (101,{-4.68});
\fill[black] (101,{-5}) rectangle (102,{-4.68});
\fill[black] (132,{-5}) rectangle (133,{-4.68});
\fill[black] (133,{-5}) rectangle (134,{-4.68});
\fill[black] (168,{-5}) rectangle (169,{-4.68});
\fill[black] (169,{-5}) rectangle (170,{-4.68});
\fill[black] (199,{-5}) rectangle (200,{-4.68});
\fill[black] (200,{-5}) rectangle (201,{-4.68});
\fill[black] (236,{-5}) rectangle (237,{-4.68});
\fill[black] (237,{-5}) rectangle (238,{-4.68});
\draw[black!40] (34,{-5}) -- (34,{-4.68});
\draw[black!40] (68,{-5}) -- (68,{-4.68});
\draw[black!40] (102,{-5}) -- (102,{-4.68});
\draw[black!40] (136,{-5}) -- (136,{-4.68});
\draw[black!40] (170,{-5}) -- (170,{-4.68});
\draw[black!40] (204,{-5}) -- (204,{-4.68});
\draw[black!40] (238,{-5}) -- (238,{-4.68});
\fill[white,opacity=0.78] (4,{-5}) rectangle (269,{-4.68});
\draw[black!55] (4,{-5.12}) -- (4,{-4.56});
\draw[thick] (4.4,{-0.1}) -- ++(0,-0.12) -- (71.6,{-0.22}) -- ++(0,0.12);
\node[below] at ({38},{-0.2}) {$\Delta_{28}\Delta_{27}$};
\draw[thick] (72.4,{-0.1}) -- ++(0,-0.12) -- (139.6,{-0.22}) -- ++(0,0.12);
\node[below] at ({106},{-0.2}) {$\Delta_{28}\Delta_{26}$};
\draw[thick] (140.4,{-0.1}) -- ++(0,-0.12) -- (207.6,{-0.22}) -- ++(0,0.12);
\node[below] at ({174},{-0.2}) {$\Delta_{28}\Delta_{25}$};
\draw[thick] (208.4,{-0.1}) -- ++(0,-0.12) -- (269,{-0.22}) ;
\node[below] at ({238.5},{-0.2}) {$\Delta_{28}\texttt{a}^{32}$};
\draw[thick] (0,{-1.47}) -- (4.6,{-1.47}) -- ++(0,0.12);
\draw[thick] (5.4,{-1.35}) -- ++(0,-0.12) -- (72.6,{-1.47}) -- ++(0,0.12);
\node[below] at ({39},{-1.45}) {$\Delta_{27}\Delta_{26}$};
\draw[thick] (73.4,{-1.35}) -- ++(0,-0.12) -- (140.6,{-1.47}) -- ++(0,0.12);
\node[below] at ({107},{-1.45}) {$\Delta_{27}\Delta_{25}$};
\draw[thick] (141.4,{-1.35}) -- ++(0,-0.12) -- (208.6,{-1.47}) -- ++(0,0.12);
\node[below] at ({175},{-1.45}) {$\Delta_{27}\Delta_{24}$};
\draw[thick] (209.4,{-1.35}) -- ++(0,-0.12) -- (269,{-1.47}) ;
\node[below] at ({239},{-1.45}) {$\Delta_{27}\texttt{a}^{32}$};
\draw[thick] (0,{-2.72}) -- (5.6,{-2.72}) -- ++(0,0.12);
\draw[thick] (6.4,{-2.6}) -- ++(0,-0.12) -- (73.6,{-2.72}) -- ++(0,0.12);
\node[below] at ({40},{-2.7}) {$\Delta_{26}\Delta_{25}$};
\draw[thick] (74.4,{-2.6}) -- ++(0,-0.12) -- (141.6,{-2.72}) -- ++(0,0.12);
\node[below] at ({108},{-2.7}) {$\Delta_{26}\Delta_{24}$};
\draw[thick] (142.4,{-2.6}) -- ++(0,-0.12) -- (209.6,{-2.72}) -- ++(0,0.12);
\node[below] at ({176},{-2.7}) {$\Delta_{26}\Delta_{23}$};
\draw[thick] (210.4,{-2.6}) -- ++(0,-0.12) -- (269,{-2.72}) ;
\node[below] at ({239.5},{-2.7}) {$\Delta_{26}\texttt{a}^{32}$};
\draw[thick] (0,{-3.97}) -- (6.6,{-3.97}) -- ++(0,0.12);
\draw[thick] (7.4,{-3.85}) -- ++(0,-0.12) -- (74.6,{-3.97}) -- ++(0,0.12);
\node[below] at ({41},{-3.95}) {$\Delta_{25}\Delta_{24}$};
\draw[thick] (75.4,{-3.85}) -- ++(0,-0.12) -- (142.6,{-3.97}) -- ++(0,0.12);
\node[below] at ({109},{-3.95}) {$\Delta_{25}\Delta_{23}$};
\draw[thick] (143.4,{-3.85}) -- ++(0,-0.12) -- (210.6,{-3.97}) -- ++(0,0.12);
\node[below] at ({177},{-3.95}) {$\Delta_{25}\Delta_{22}$};
\draw[thick] (211.4,{-3.85}) -- ++(0,-0.12) -- (269,{-3.97}) ;
\node[below] at ({240},{-3.95}) {$\Delta_{25}\texttt{a}^{28}$};
\draw[thick] (0,{-5.22}) -- (3.6,{-5.22}) -- ++(0,0.12);
\fill[orange!85!black] (4.5,{0.34}) -- ++(-1.6,0.22) -- ++(3.2,0) -- cycle;
\fill[orange!85!black] (5.5,{-0.91}) -- ++(-1.6,0.22) -- ++(3.2,0) -- cycle;
\fill[orange!85!black] (6.5,{-2.16}) -- ++(-1.6,0.22) -- ++(3.2,0) -- cycle;
\fill[orange!85!black] (7.5,{-3.41}) -- ++(-1.6,0.22) -- ++(3.2,0) -- cycle;
\end{tikzpicture}
\caption{Setting of \cref{lem:rows}: the rows of the factorization of the hard suffix of~$B_{32}$ ($k = 32$, $m = 4$, $|V_{32,4}| = 269$), which is the same for LZD, LZD+, and LZDR. 
Each strip is one copy of~$V_{32,4}$ as it occurs in $W = V_{32,4}^\infty$, with the letters~$\texttt{b}$ in black, and the labels on top name its blocks.
Since the hard suffix is $W[m+1..m+m|V_{32,4}|]$, it begins $m = 4$ letters inside copy~$1$, after the gray cells, and ends $m$ letters inside the following copy; in the last strip, labeled \emph{tail}, only the $m$ letters left of the vertical line belong to the text, namely the letters that the rotation moves from the front to the end of each copy of~$V'_{32,4}$; the faded part lies behind the end of~$B_{32}$.
Head and tail together therefore amount to one more copy of~$V_{32,4}$, so the $m$ copies of~$V'_{32,4}$ span $m+1$ strips. 
The brackets below a strip are the factors, labeled by their two components; the last factor of each row continues in the next strip. 
Each row starts one letter further inside its copy than the previous row (down-pointing orange triangles), so relative to the dashed boundaries of the pairs $\Delta_k\Delta_{k-j}$, the factor boundaries move one letter to the right per copy, and a row starting at phase~$r$ reads each block~$\Delta_i$ as~$\Delta_{i-r}$.}
\label{fig:rows}

\end{figure}

\begin{figure}[t]

\centering
\begin{tikzpicture}[x=0.21961cm,y=1cm,font=\scriptsize]
\node[anchor=west] at (0,{1.03}) {(a) without rotation, at phase~$r = 1$};
\draw[fill=white,draw=black!55] (0,{0}) rectangle (15,{0.4});
\node at ({7.5},{0.2}) {$\texttt{a}^{15}$};
\draw[fill=black!22,draw=black!55] (15,{0}) rectangle (17,{0.4});
\node at ({16},{0.2}) {$\texttt{b}^{2}$};
\draw[fill=white,draw=black!55] (17,{0}) rectangle (32,{0.4});
\node at ({24.5},{0.2}) {$\texttt{a}^{15}$};
\draw[fill=black!22,draw=black!55] (32,{0}) rectangle (34,{0.4});
\node at ({33},{0.2}) {$\texttt{b}^{2}$};
\draw[fill=white,draw=black!55] (34,{0}) rectangle (51,{0.4});
\node at ({42.5},{0.2}) {$\texttt{a}^{17}$};
\draw[thick] (0.15,{0.48}) -- ++(0,0.12) -- (17.85,{0.6}) -- ++(0,-0.12);
\node[above] at ({9},{0.55}) {$\Delta_{15}$};
\draw[thick] (18.15,{0.48}) -- ++(0,0.12) -- (35.85,{0.6}) -- ++(0,-0.12);
\node[above] at ({27},{0.55}) {$\Delta_{14}$};
\node[anchor=west] at (37,{0.58}) {combination: $2k+4 = 36$};
\draw[thick,orange!85!black] (0.15,{-0.08}) -- ++(0,-0.12) -- (16.85,{-0.2}) -- ++(0,0.12);
\node[below] at ({8.5},{-0.15}) {$\Gamma_{15}$};
\draw[thick,orange!85!black] (17.15,{-0.08}) -- ++(0,-0.12) -- (33.85,{-0.2}) -- ++(0,0.12);
\node[below] at ({25.5},{-0.15}) {$\Gamma_{15}$};
\draw[thick,orange!85!black] (34.15,{-0.08}) -- ++(0,-0.12) -- (48.85,{-0.2}) -- ++(0,0.12);
\node[below] at ({41.5},{-0.15}) {$\texttt{a}^{15}$};
\draw[orange!85!black,densely dotted] (49,{-0.24}) -- (49,{0.4});
\node[anchor=west] at (0,{-0.62}) {repetition of $\Gamma_{15}$: $3p+4 = 49$, so the next factor is the repetition};
\node[anchor=west] at (0,{-1.42}) {(b) in~$B_{16}$, at phase~$r = m = 2$};
\draw[fill=white,draw=black!55] (0,{-2.45}) rectangle (14,{-2.05});
\node at ({7},{-2.25}) {$\texttt{a}^{14}$};
\draw[fill=black!22,draw=black!55] (14,{-2.45}) rectangle (16,{-2.05});
\node at ({15},{-2.25}) {$\texttt{b}^{2}$};
\draw[fill=white,draw=black!55] (16,{-2.45}) rectangle (31,{-2.05});
\node at ({23.5},{-2.25}) {$\texttt{a}^{15}$};
\draw[fill=black!22,draw=black!55] (31,{-2.45}) rectangle (33,{-2.05});
\node at ({32},{-2.25}) {$\texttt{b}^{2}$};
\draw[fill=white,draw=black!55] (33,{-2.45}) rectangle (50,{-2.05});
\node at ({41.5},{-2.25}) {$\texttt{a}^{17}$};
\draw[thick] (0.15,{-1.97}) -- ++(0,0.12) -- (17.85,{-1.85}) -- ++(0,-0.12);
\node[above] at ({9},{-1.9}) {$\Delta_{14}$};
\draw[thick] (18.15,{-1.97}) -- ++(0,0.12) -- (35.85,{-1.85}) -- ++(0,-0.12);
\node[above] at ({27},{-1.9}) {$\Delta_{13}$};
\node[anchor=west] at (37,{-1.87}) {combination: $2k+4 = 36$};
\draw[thick,orange!85!black] (0.15,{-2.53}) -- ++(0,-0.12) -- (15.85,{-2.65}) -- ++(0,0.12);
\node[below] at ({8},{-2.6}) {$\Gamma_{14}$};
\draw[thick,orange!85!black] (16.15,{-2.53}) -- ++(0,-0.12) -- (29.85,{-2.65}) -- ++(0,0.12);
\node[below] at ({23},{-2.6}) {$\texttt{a}^{14}$};
\draw[orange!85!black,densely dotted] (30,{-2.69}) -- (30,{-2.05});
\node[anchor=west] at (0,{-3.07}) {repetition of $\Gamma_{14}$: $2p+2 = 30$, so the next factor is the combination};
\end{tikzpicture}
\caption{Why the hard part is rotated (\cref{lem:rows}), shown for $k = 16$, where $m = 2$; each box is a run of equal letters, and the runs of~$\texttt{b}$ are shaded.
At the $j$th factor of a row starting at phase~$r$, the text begins with $\Delta_p\Delta_{p-j}$ with $p = k-r$, so the combination (above the text) has length $2k+4$.
Since the schemes select a longest candidate, the factor is the combination unless some other candidate is longer.
Every other candidate is shorter than $2k+4$, except possibly the repetition of $\Gamma_p = \texttt{a}^p\texttt{bb}$ (below the text, orange), so it decides the factor.
(a)~For $j = r$, the second copy of~$\Gamma_p$ matches completely, so this repetition reaches $3p+4 > 2k+4$ letters and is selected; this happens at phase~$1$ of an unrotated hard suffix, and it is how LZDR escapes the rows on~$S_k$.
(b)~In~$B_k$, every row starts at a phase $r \ge m > j$, so the second copy breaks off after $p$ letters~$\texttt{a}$, the repetition reaches only $2p+2 < 2k+4$ letters, and the combination is selected.}
\label{fig:candidates}

\end{figure}

\begin{proof}[Proof of \cref{thm:binary}]

By \cref{lem:binary-prefix,lem:rows}, each of the three schemes produces the same $7k/2-3+m^2 = 7k/2-3+k^2/64$ factors.
Moreover, $|B_k| = |U_k| + m|V_{k,m}| = O(k^2) + (k/8)\bigl((k+2)k/4-3\bigr) = \Theta(k^3)$.
An SLP of size $O(k)$ generates~$U_k$ by \cref{lem:binary-prefix}, $V_{k,m}$ with $O(k)$ further rules, $V'_{k,m} = V_{k,m}[m+1..|V_{k,m}|]\,V_{k,m}[1..m]$ from the same blocks together with $\texttt{a}^{k-m}$ and~$\texttt{a}^m$, and $(V'_{k,m})^m$ with $O(\log m)$ doubling rules.
Hence, $g(B_k) = O(k)$, and $z(B_k)/g(B_k) = \Omega(k) = \Omega(|B_k|^{1/3})$ for each of the three schemes.

\end{proof}

\subsection{Upper bounds}\label{sec:bounds-upper}

The upper bound counts distinct substrings, for which we first observe that factors do not repeat: intuitively, a factor that equals an earlier factor could have been extended by one letter, contradicting the greedy choice.

\begin{lemma}\label[lemma]{lem:distinct}

All factors of the greedy LZD+ factorization, except possibly the last one, are pairwise distinct.
The same holds for LZDR and, as shown by Goto et al.~\cite{goto15lzd} (see also \cite[Lemma~3]{badkobeh17two}), for LZD.

\end{lemma}
\begin{proof}

Suppose that $F_i = F_j$ for some $j < i < z$.
When $F_i$ is computed, $F_j$ is a previous factor matching the text, so the first component~$G$ of the combination rule, i.e., the longest complete reference, satisfies $|G| \ge |F_i|$.
If $|G| > |F_i|$, then $G[1..|F_i|+1]$ is a truncation candidate (for LZDR, a repetition candidate) longer than~$F_i$.
If $|G| = |F_i|$, then $G$ followed by the next letter, which exists since $i < z$, is a combination candidate of length $|F_i|+1$.
Both cases contradict the greedy choice of~$F_i$.

\end{proof}

We adapt the counting argument of
Charikar et al.~\cite{charikar05grammar} for LZ78,
subsequently used by
Badkobeh et al.~\cite[Theorem~5]{badkobeh17two} for LZD.
Let $T$ be a text of length~$n$, and let
$\delta = \max_{1 \le \ell \le n} d_\ell(T)/\ell$ denote the \emph{substring complexity} of $T$,
where $d_\ell(T)$ denotes the number of distinct substrings of~$T$ of length~$\ell$.
Then the following holds.

\begin{theorem}\label[theorem]{thm:attractor}
Each of $z_{\mathrm{LZD}}$, $z_+$, and $z_R$
is in
\(O(n^{2/3}\delta^{1/3}) \subseteq O(n^{2/3}g^{1/3}).\)
\end{theorem}
\begin{proof}
By the definition of~$\delta$, the text~$T$
has at most $\delta\ell$ distinct substrings
of length~$\ell$.
By \cref{lem:distinct}, all factors except
possibly the last one are pairwise distinct.
Thus, for every integer threshold $\lambda \ge 1$,
the number of factors of length at most~$\lambda$,
excluding the last factor, is bounded by
\[
\sum_{\ell=1}^{\lambda} d_\ell(T)
\le \delta\sum_{\ell=1}^{\lambda}\ell
= \frac{\delta\lambda(\lambda+1)}{2}
\le \delta\lambda^2.
\]
Since the factors form a partition of~$T$,
at most $n/\lambda$ factors have length
greater than~$\lambda$.
Consequently, the total number of factors is at most
\(
\delta\lambda^2 + \frac{n}{\lambda} + 1.
\)
Choosing
$\lambda := \lceil(n/\delta)^{1/3}\rceil$
balances the first two terms and yields
$O(n^{2/3}\delta^{1/3})$ factors.

Finally, $\delta \le 2g$: a grammar whose right-hand sides have total length~$m$ has at most $m\ell$ distinct substrings of length~$\ell$~\cite{charikar05grammar} (see also \cite[Lemma~2]{badkobeh17two}), since each of them is determined by a symbol on a right-hand side and an offset in $[0..\ell-1]$; the right-hand sides of an SLP with $g$ rules have total length at most~$2g$.
The claimed bounds follow.
\end{proof}

For LZDR, there is another bound in terms of run-length encoding.
A \emph{letter run} of~$T$ is a maximal substring of~$T$ that consists of a single repeated letter; let $\operatorname{runs}(T)$ denote the number of letter runs of~$T$, and write $\operatorname{runs}$ if $T$ is clear from the context.

\begin{theorem}\label[theorem]{thm:run}

For every text~$T$, $z_R \le \operatorname{runs}(T)$.

\end{theorem}
\begin{proof}

Suppose that a factor~$F_x$ starts at a position from which its letter run continues for $e$ letters, counting the starting position.
If $e \ge 2$, the repetition rule with this letter yields a candidate of length~$e$; if $e = 1$, the factor has length at least one anyway.
In both cases, $F_x$ ends at or after the end of this run, so no two factors start in the same run.

\end{proof}

This bound is tight up to constant factors on the strings~$B_k$ of \cref{thm:binary}: each copy of~$V'_{k,m}$ has $\Theta(m)$ letter runs, so $\operatorname{runs}(B_k) = \Theta(k + m^2) = \Theta(k^2) = \Theta(z_R(B_k))$.
In total, $z_R = O(\min\{\operatorname{runs}, n^{2/3}\delta^{1/3}\})$, and by \cref{thm:binary,thm:attractor}, the approximation ratio of each of LZD, LZD+, and LZDR lies between $\Omega(n^{1/3})$ and $O((n/\log n)^{2/3})$.

\section{Collage Systems}\label{sec:collage}

A \emph{collage system} is a framework for compressed pattern matching that can be used to represent a string $T$ as a pair of a dictionary $D$ and a sequence of phrases $S$~\cite{kida03collage}.
The dictionary $D$ is a sequence of $d$ assignments $X_k = expr_k$ for $k \in [1..d]$ that represent phrases, where $X_k$ is a token and $expr_k$ is one of the five expression forms:
\begin{itemize}
    \item Primitive: $X_k = a$ for $a \in \Sigma \cup \{ \varepsilon \}$
    \item Concatenation: $X_k = X_i X_j$ for $i, j < k$
    \item Prefix truncation: $X_k = X_i [j..]$ for $i < k$ and $1 \leq j \leq |X_i|$
    \item Suffix truncation: $X_k = X_i [1..j]$ for $i < k$ and  $1 \leq j \leq |X_i|$
    \item Repetition: $X_k = (X_i)^j$ for $i < k$ and $j \in \mathbb{N}^+$
\end{itemize}
$S$ is used to concatenate phrases from the dictionary such that it represents the string $T$.
The size of $D$ is the number of assignments $d$ and is denoted by $\left\lVert D \right\rVert$.
The size of $S = X_{i_1} X_{i_2} \cdots X_{i_k}$ is the number of phrases $k$ used in $S$, and is denoted by $|S|$.
We define the size of a collage system as $\left\lVert D \right\rVert + |S|$.

It is possible to represent LZD+ as a collage system that generates the string $T$ using $\Theta(z)$ assignments.
This is because each factor can be expressed as a constant number of assignments, such that each factor representation only needs a maximum of two primitive assignments, and a concatenation and/or suffix truncation assignment.

It is also possible to represent LZDR as a collage system that generates the string $T$ using $\Theta(z)$ assignments. This is because each factor can be expressed as a constant number of assignments, such that each factor representation only needs a maximum of two primitive assignments, maybe a concatenation or repetition assignment, and finally maybe a suffix truncation assignment.

These representations are induced by the factorizations; they do not imply that a smallest collage system for~$T$ has $\Theta(z)$ assignments.
The same construction applies to \AltFlex{}-LZDR, but not directly to \StdFlex{}-LZDR and \MaxAlt{}-LZDR, whose factors reference greedy factors that are not part of their output.

We can also consider the lower bound example for LZD in \cref{sec:lowerbound} with respect to a collage system.
Since every factor in LZDR can be expressed with $O(1)$ assignments in a collage system, the size of $D$ would be $\Theta(k)$ if the factor count $6k+3$ observed in \cref{tab:lowerBound} held for every~$k$, and the sequence $S$ for $S_k$ would then be the concatenation of the tokens of the $6k+3$ factors, i.e., $S = X_{i_1} X_{i_2} \cdots X_{i_{6k+3}}$, such that $|S| = \Theta(k)$.
Therefore, the size of the collage system would be $\Theta(k)$, yielding the same size upper bound as the size of the smallest grammar of $O(k)$.

On the binary strings~$B_k$ of \cref{thm:binary}, in contrast, the collage system induced by LZDR has $\Omega(k^2)$ assignments, although an SLP, and hence a collage system, of size $O(k)$ exists.

\begin{table}[t]
  \scriptsize
  \begin{tabular}{|l|r|r|r|r|r|r|r|r|}
        \hline
    {} &
    {$n$} &
		\multicolumn{4}{c|}{LZDR}&
    {} &
    {} &
		{}
		\\
    {dataset} & [MiB] &
    {greedy} &
		{\StdFlex{} } &
		{\AltFlex{} } &
		{\MaxAlt{}} &
    {LZD+} &
    {LZ78} &
    {LZW}
        \\\hline
\textsc{sources } & 50   & $1.28\%$ & $-7.58\%$ & $\mathbf{-8.05\%}$ & $-3.83\%$ & $1.30\%$ & $99.48\%$ & $124.46\%$ \\\hline
\textsc{sources } & 100  & $1.35\%$ & $-8.02\%$ & $\mathbf{-8.52\%}$ & $-4.02\%$ & $1.37\%$ & $98.65\%$ & $122.72\%$ \\\hline
\textsc{sources } & 200  & $1.83\%$ & $-7.89\%$ & $\mathbf{-8.58\%}$ & $-3.72\%$ & $1.87\%$ & $102.34\%$ & $124.96\%$ \\\hline
\textsc{pitches } & 50   & $-4.40\%$ & $-10.22\%$ & $\mathbf{-11.92\%}$ & $-7.37\%$ & $-3.88\%$ & $57.91\%$ & $89.20\%$ \\\hline
\textsc{proteins } &50  & $-0.38\%$ & $-5.76\%$ & $\mathbf{-8.40\%}$ & $-2.68\%$ & $-0.38\%$ & $49.68\%$ & $77.54\%$ \\\hline
\textsc{proteins } &100 & $-0.22\%$ & $-6.51\%$ & $\mathbf{-9.71\%}$ & $-3.37\%$ & $-0.25\%$ & $53.53\%$ & $80.87\%$ \\\hline
\textsc{proteins } &200 & $-0.46\%$ & $-7.30\%$ & $\mathbf{-11.06\%}$ & $-3.51\%$ & $-0.49\%$ & $61.64\%$ & $89.03\%$ \\\hline
\textsc{dna } & 50       & $-2.79\%$ & $\mathbf{-13.78\%}$ & $-12.16\%$ & $-7.02\%$ & $-3.30\%$ & $39.58\%$ & $52.32\%$ \\\hline
\textsc{dna } & 100      & $-2.55\%$ & $\mathbf{-13.92\%}$ & $-12.28\%$ & $-6.95\%$ & $-3.09\%$ & $39.53\%$ & $51.74\%$ \\\hline
\textsc{dna } & 200      & $-2.35\%$ & $\mathbf{-14.10\%}$ & $-12.43\%$ & $-6.88\%$ & $-2.90\%$ & $40.02\%$ & $51.81\%$ \\\hline
\textsc{english } & 50   & $0.62\%$ & $-6.08\%$ & $\mathbf{-6.79\%}$ & $-1.23\%$ & $0.67\%$ & $86.54\%$ & $107.55\%$ \\\hline
\textsc{english } & 100  & $1.04\%$ & $-6.06\%$ & $\mathbf{-6.68\%}$ & $-1.38\%$ & $1.05\%$ & $82.00\%$ & $101.37\%$ \\\hline
\textsc{english } & 200  & $1.62\%$ & $-5.74\%$ & $\mathbf{-6.57\%}$ & $-0.73\%$ & $1.59\%$ & $84.80\%$ & $103.09\%$ \\\hline
\textsc{xml } & 50  & $0.68\%$ & $\mathbf{-7.47\%}$ & $-6.47\%$ & $-3.45\%$ & $0.66\%$ & $98.68\%$ & $125.63\%$ \\\hline
\textsc{xml } & 100 & $1.41\%$ & $\mathbf{-7.34\%}$ & n/a & $-3.05\%$ & $1.34\%$ & $103.05\%$ & $128.52\%$ \\\hline
\textsc{xml } & 200 & $2.51\%$ & $\mathbf{-6.80\%}$ & n/a & $-2.17\%$ & $2.45\%$ & $109.91\%$ & $134.72\%$ \\\hline
\end{tabular}
  \caption{Pizza\&Chili corpus factor counts relative to LZD\@.
      The second column denotes the prefix in MiB we extracted from the respective dataset.
      We marked the fewest factor counts in bold.
  We have two entries with \emph{n/a} for which the computation did not finish within several hours.}
  \label{tab:pizzachili}
\end{table}

\section{Practical benchmarks}

In what follows, we compare the factor count, execution time, and maximum memory usage across three different corpora: 
the Calgary corpus~\cite{bell89modeling}, the Canterbury corpus~\cite{arnold97corpus}, and the Pizza\&Chili corpus~\cite{ferragina08compressed} for large files. The Calgary file \texttt{pic} and the Canterbury file \texttt{ptt5} are the same file under different names; \cref{tab:calgary} therefore lists it only once, as \texttt{pic}.
We compared the following compression schemes: 
LZDR, its flexible parsing variants, LZD+, LZD~\cite{goto15lzd}, LZ78~\cite{ziv78lz}, LZW~\cite{welch84lzw}, and the flexible parsing variants of LZW as defined by Matias et al.~\cite{matias99optimality}, which we here call \StdFlex{}-LZW, and by Horspool~\cite{horspool95effect}, which we here call \AltFlex{}-LZW\@.\footnote{We did not include LZ-ABT~\cite{ohno18lzabt} since it seems that the available implementation (\url{https://github.com/tatsuya0619/lzabt}) does not report factor counts, and reimplementing the proposed factorization scheme seems rather tricky compared to LZD\@.
}
To improve distinguishability, we tagged the standard (i.e., greedy) LZDR and LZW parsings with \emph{greedy} in the experiments.
We compared the execution time and maximum memory usage for a subset of these, namely: LZDR, LZD+, LZD, LZ78, and LZW\@.

Our experiments measure numbers of factors, not compressed sizes in bits.
An LZD factor consists of two references, whereas an LZD+ or LZDR factor additionally has to specify its production rule and the length~$\ell_x$, which can cost up to $\lceil \lg n \rceil$ additional bits.
Hence, fewer factors do not by themselves imply a smaller compressed size; we leave the encoding as an open problem (\cref{sec:conclusion}).

\block{Source code and setup}
For LZD, we used the already existing implementation described in~\cite{goto15lzd}\footnote{\url{https://github.com/kg86/lzd} at commit 79498a5}.
The existing LZD implementation did not output our expected factor count number, however, so while we used the existing implementation for determining the execution time and maximum memory usage, we also implemented our own LZD parsing to determine the factor count. Hence, for LZD, the reported running times and the reported factor counts stem from different implementations, whose factorizations may differ.
To determine the factor count, execution time, and maximum memory usage for LZ78 and LZW, we used tudocomp~\cite{dinklage17tudocomp}\footnote{\url{https://github.com/tudocomp/tudocomp} at commit b5512f85, with parameters \texttt{coder=ascii} and the defaults \texttt{trie=ternary} and unlimited dictionary size}.
All other compression schemes have been implemented on our own in C++.
Our code is freely accessible on GitHub at \url{https://github.com/koeppl/lzdr-comp}.
All programs have been compiled with the flags \texttt{-O3 -DNDEBUG}.
We ran our benchmarks on an Intel Core i5-4590 with 32 GB RAM\@.
The execution time was determined by using the median execution time of 5 separate runs.

\block{Implementation details}
While our proposed linear-time LZDR algorithm works with an LCE data structure, we were faster with letter-wise comparisons in practice compared to practical LCE data structures\footnote{\url{https://github.com/herlez/lce-test} at commit b52e00a using SSS512}. 
For this reason we opted to implement $T.\fnLCE$ naively.
By calling each helper function individually, we need to perform multiple times a trie traversal following the same path.
In practice, the hash table lookups and the node traversals are costly.
That is why we collapse for LZD+ and LZDR the helper functions such that we need two trie traversals for determining each $F_{x_1}$ and $F_{x_2}$ for $F_x$, during which we also compute the truncation and repetition rules.

\begin{figure}[t]
\centering
\begin{tikzpicture}
  \begin{groupplot}[
      group style={
          group size=3 by 2, horizontal sep=2cm, vertical sep=2cm, },
      xlabel={Input size (MiB)},
      ylabel={Time (seconds)},
      ymin=0, ymax=60,
      xmin=0, xmax=250,
      xtick={50,100,200},
      width=0.3\textwidth, height=3.5cm, grid=major
  ]

\nextgroupplot[
      title={\textsc{dna}},
      legend style = { column sep = 6pt, legend columns = 1, legend to name = group legend time },
      legend cell align={left}
  ]

\addplot[
      only marks,
      mark=triangle,
      color=blue
  ] coordinates {
      (100, 26.7)
      (200, 55.3)
      (50, 11.7)
  };
  \addlegendentry{LZDR}

\addplot[
      only marks,
      mark=square,
      color=red
  ] coordinates {
      (100, 25.7)
      (200, 54.3)
      (50, 11.4)
  };
  \addlegendentry{LZD+}

\addplot[
      only marks,
      mark=o,
      color=green
  ] coordinates {
      (100, 23.1)
      (200, 50.0)
      (50, 10.1)
  };
  \addlegendentry{LZD}

\addplot[
      only marks,
      mark=star,
      color=purple
  ] coordinates {
      (100, 10.3)
      (200, 22.3)
      (50, 4.6)
  };
  \addlegendentry{LZ78}

\addplot[
      only marks,
      mark=x,
      color=orange
  ] coordinates {
      (100, 10.8)
      (200, 23.7)
      (50, 4.9)
  };
  \addlegendentry{LZW}

\nextgroupplot[
      title={\textsc{english}}
  ]

\addplot[
      only marks,
      mark=triangle,
      color=blue
  ] coordinates {
      (100, 21.1)
      (200, 48.4)
      (50, 9.6)
  };

\addplot[
      only marks,
      mark=square,
      color=red
  ] coordinates {
      (100, 20.7)
      (200, 46.9)
      (50, 9.5)
  };

\addplot[
      only marks,
      mark=o,
      color=green
  ] coordinates {
      (100, 21.2)
      (200, 46.3)
      (50, 9.8)
  };

\addplot[
      only marks,
      mark=star,
      color=purple
  ] coordinates {
      (100, 14.6)
      (200, 31.4)
      (50, 6.7)
  };

\addplot[
      only marks,
      mark=x,
      color=orange
  ] coordinates {
      (100, 15.7)
      (200, 33.6)
      (50, 7.2)
  };

\nextgroupplot[
      title={\textsc{proteins}}
  ]

\addplot[
      only marks,
      mark=triangle,
      color=blue
  ] coordinates {
      (100, 26.3)
      (200, 51.5)
      (50, 12.1)
  };

\addplot[
      only marks,
      mark=square,
      color=red
  ] coordinates {
      (100, 25.5)
      (200, 50.3)
      (50, 11.8)
  };

\addplot[
      only marks,
      mark=o,
      color=green
  ] coordinates {
      (100, 30.7)
      (200, 59.2)
      (50, 13.5)
  };

\addplot[
      only marks,
      mark=star,
      color=purple
  ] coordinates {
      (100, 19.6)
      (200, 41.3)
      (50, 8.8)
  };

\addplot[
      only marks,
      mark=x,
      color=orange
  ] coordinates {
      (100, 22.7)
      (200, 46.9)
      (50, 10.3)
  };

\nextgroupplot[
      title={\textsc{sources}}
  ]

\addplot[
      only marks,
      mark=triangle,
      color=blue
  ] coordinates {
      (100, 13.5)
      (200, 28.8)
      (50, 6.2)
  };

\addplot[
      only marks,
      mark=square,
      color=red
  ] coordinates {
      (100, 13.2)
      (200, 27.9)
      (50, 6.1)
  };

\addplot[
      only marks,
      mark=o,
      color=green
  ] coordinates {
      (100, 15.0)
      (200, 32.9)
      (50, 6.6)
  };

\addplot[
      only marks,
      mark=star,
      color=purple
  ] coordinates {
      (100, 11.5)
      (200, 24.9)
      (50, 5.4)
  };

\addplot[
      only marks,
      mark=x,
      color=orange
  ] coordinates {
      (100, 12.0)
      (200, 26.3)
      (50, 5.7)
  };

\nextgroupplot[
      title={\textsc{xml}}
  ]

\addplot[
      only marks,
      mark=triangle,
      color=blue
  ] coordinates {
      (100, 8.0)
      (200, 17.5)
      (50, 3.8)
  };

\addplot[
      only marks,
      mark=square,
      color=red
  ] coordinates {
      (100, 7.9)
      (200, 17.3)
      (50, 3.7)
  };

\addplot[
      only marks,
      mark=o,
      color=green
  ] coordinates {
      (100, 8.6)
      (200, 19.3)
      (50, 4.1)
  };

\addplot[
      only marks,
      mark=star,
      color=purple
  ] coordinates {
      (100, 7.3)
      (200, 16.4)
      (50, 3.3)
  };

\addplot[
      only marks,
      mark=x,
      color=orange
  ] coordinates {
      (100, 7.9)
      (200, 17.9)
      (50, 3.6)
  };
  \end{groupplot}

\coordinate (xmid) at ($(group c3r1.north east)!0.5!(group c3r1.north west)$);
  \coordinate (ybot) at ($(current bounding box.south)!0.5!(group c3r1.south west)$);
  \coordinate (legpos) at (xmid |- ybot);
  \node at (legpos) [yshift=-0.5cm] {\pgfplotslegendfromname{group legend time}};
\end{tikzpicture}
\caption{Execution time in relation to input size for multiple compression schemes}
\label{fig:exectime}
\end{figure}

\subsection{Factor count}
We analyze the factor count on the two corpora with small file sizes and the Pizza\&Chili corpus separately.

\block{Calgary and Canterbury dataset}
From \cref{tab:calgary} we observe that
LZDR, the flexible parsing variants of LZDR, as well as LZD+ almost consistently achieve a lower factor count than LZD\@.
LZDR and LZD+ have a relatively similar improvement over LZD, while the LZDR flexible parsing variants show the best improvement.
\StdFlex{}-LZDR achieved the lowest factor count of the evaluated compression schemes the most times, and \MaxAlt{}-LZDR performed worst on average of all LZDR flexible parsing variants.
In one case, \StdFlex{}-LZDR even attained a {\raise.17ex\hbox{$\scriptstyle\mathtt{\sim}$}}10\% improvement relative to LZD\@.
LZW on the other hand has consistently the highest factor count of all.
While the LZW flexible parsing variants achieve a smaller factor count than LZW, the results are still inferior to LZ78.
Nonetheless, LZ78 shows a higher factor count than LZD.

\begin{figure}[t]
\centering
\begin{tikzpicture}
  \begin{groupplot}[
    group style={
      group size=3 by 2, horizontal sep=2cm, vertical sep=2cm, },
    xlabel={Input size (MiB)},
    ylabel={Memory (MiB)},
    ymin=0, ymax=4096,
    xmin=0, xmax=250,
    xtick={50,100,200},
    ytick={1024,2048,3072,4096},
    width=0.3\textwidth, height=3.5cm, grid=major
  ]

\nextgroupplot[
      title={\textsc{dna}},
      legend style = { column sep = 10pt, legend columns = 1, legend to name = group legend mem },
      legend cell align={left}
  ]

\addplot[
      only marks,
      mark=triangle,
      color=blue
  ] coordinates {
      (100, 1536.0)
      (200, 2969.6)
      (50, 817.8)
  };
  \addlegendentry{LZDR}

\addplot[
      only marks,
      mark=square,
      color=red
  ] coordinates {
      (100, 1536.0)
      (200, 2969.6)
      (50, 813.8)
  };
  \addlegendentry{LZD+}

\addplot[
      only marks,
      mark=o,
      color=green
  ] coordinates {
      (100, 1228.8)
      (200, 2355.2)
      (50, 661.3)
  };
  \addlegendentry{LZD}

\addplot[
      only marks,
      mark=star,
      color=purple
  ] coordinates {
      (100, 172.3)
      (200, 306.3)
      (50, 102.8)
  };
  \addlegendentry{LZ78}

\addplot[
      only marks,
      mark=x,
      color=orange
  ] coordinates {
      (100, 162.2)
      (200, 296.0)
      (50, 111.5)
  };
  \addlegendentry{LZW}

\nextgroupplot[
      title={\textsc{english}}
  ]

\addplot[
      only marks,
      mark=triangle,
      color=blue
  ] coordinates {
      (100, 1433.6)
      (200, 2662.4)
      (50, 712.8)
  };

\addplot[
      only marks,
      mark=square,
      color=red
  ] coordinates {
      (100, 1433.6)
      (200, 2662.4)
      (50, 712.8)
  };

\addplot[
      only marks,
      mark=o,
      color=green
  ] coordinates {
      (100, 1024.0)
      (200, 1945.6)
      (50, 551.4)
  };

\addplot[
      only marks,
      mark=star,
      color=purple
  ] coordinates {
      (100, 206.0)
      (200, 370.4)
      (50, 121.0)
  };

\addplot[
      only marks,
      mark=x,
      color=orange
  ] coordinates {
      (100, 228.7)
      (200, 411.4)
      (50, 133.2)
  };

\nextgroupplot[
      title={\textsc{proteins}}
  ]

\addplot[
      only marks,
      mark=triangle,
      color=blue
  ] coordinates {
      (100, 2048.0)
      (200, 3584.0)
      (50, 1024.0)
  };

\addplot[
      only marks,
      mark=square,
      color=red
  ] coordinates {
      (100, 2048.0)
      (200, 3584.0)
      (50, 1024.0)
  };

\addplot[
      only marks,
      mark=o,
      color=green
  ] coordinates {
      (100, 1740.8)
      (200, 3276.8)
      (50, 941.8)
  };

\addplot[
      only marks,
      mark=star,
      color=purple
  ] coordinates {
      (100, 299.0)
      (200, 572.6)
      (50, 374.5)
  };

\addplot[
      only marks,
      mark=x,
      color=orange
  ] coordinates {
      (100, 343.1)
      (200, 611.9)
      (50, 325.6)
  };

\nextgroupplot[
      title={\textsc{sources}}
  ]

\addplot[
      only marks,
      mark=triangle,
      color=blue
  ] coordinates {
      (100, 1228.8)
      (200, 2252.8)
      (50, 647.8)
  };

\addplot[
      only marks,
      mark=square,
      color=red
  ] coordinates {
      (100, 1228.8)
      (200, 2252.8)
      (50, 648.8)
  };

\addplot[
      only marks,
      mark=o,
      color=green
  ] coordinates {
      (100, 921.0)
      (200, 1740.8)
      (50, 509.0)
  };

\addplot[
      only marks,
      mark=star,
      color=purple
  ] coordinates {
      (100, 203.4)
      (200, 361.1)
      (50, 121.1)
  };

\addplot[
      only marks,
      mark=x,
      color=orange
  ] coordinates {
      (100, 227.4)
      (200, 401.0)
      (50, 133.8)
  };

\nextgroupplot[
      title={\textsc{xml}}
  ]

\addplot[
      only marks,
      mark=triangle,
      color=blue
  ] coordinates {
      (100, 727.8)
      (200, 1433.6)
      (50, 386.8)
  };

\addplot[
      only marks,
      mark=square,
      color=red
  ] coordinates {
      (100, 727.8)
      (200, 1433.6)
      (50, 386.8)
  };

\addplot[
      only marks,
      mark=o,
      color=green
  ] coordinates {
      (100, 620.7)
      (200, 1126.4)
      (50, 333.1)
  };

\addplot[
      only marks,
      mark=star,
      color=purple
  ] coordinates {
      (100, 123.2)
      (200, 212.4)
      (50, 78.3)
  };

\addplot[
      only marks,
      mark=x,
      color=orange
  ] coordinates {
      (100, 136.5)
      (200, 237.5)
      (50, 85.8)
  };
  \end{groupplot}

  \coordinate (xmid) at ($(group c3r1.north east)!0.5!(group c3r1.north west)$);
  \coordinate (ybot) at ($(current bounding box.south)!0.5!(group c3r1.south west)$);
  \coordinate (legpos) at (xmid |- ybot);
  \node at (legpos) [yshift=-0.7cm] {\pgfplotslegendfromname{group legend mem}};
\end{tikzpicture}
\caption{Maximum memory usage in relation to input size for multiple compression schemes}
\label{fig:maxmem}
\end{figure}

\block{Pizza\&Chili corpus}
For larger files of the Pizza\&Chili corpus, the results are given in \cref{tab:pizzachili}.
For most of the datasets, we create files by extracting prefixes of lengths 50, 100, and 200 MiB.
Our observations are mostly the same, but with some differences:
\begin{enumerate}
  \item The improvement of LZDR and LZD+ over LZD disappears; these three compression schemes now perform about the same.
  \item The flexible parsing variants of LZDR still consistently achieve the best results. This time, \AltFlex{}-LZDR attains the smallest factor count for most datasets (\textsc{sources}, \textsc{pitches}, \textsc{proteins}, and \textsc{english}),
      but \StdFlex{}-LZDR still holds up great and even achieved the highest improvement for a single file of around 14\%.
  \item Only \MaxAlt{}-LZDR performs poorly in comparison to the two previous flexible parsing LZDR variants, though still an improvement compared to LZD.
\end{enumerate}

At a fixed position and with the same dictionary, every LZD factor is an LZD+ candidate, and every LZD+ factor is an LZDR candidate.
This does not imply that LZD+ and LZDR produce fewer factors than LZD, since a longer factor shifts all later factor boundaries and thus changes the dictionary.
\Cref{tab:pizzachili} shows this effect: on \textsc{sources}, \textsc{english}, and \textsc{xml}, greedy LZD+ and LZDR produce up to $2.51\%$ more factors than LZD.

While the Horspool variants for LZW in~\cite[Table~4]{matias01effect} as well as for LZ78 in~\cite[Table~3]{DBLP:journals/corr/abs-2409-14649} have empirically fewer factors than the standard flexible parsing for most datasets,
our experiments reveal that the same statement cannot be made when comparing \StdFlex{}-LZDR with \MaxAlt{}-LZDR.

\newcommand{\Dataset}[1]{\textsc{#1}}

\begin{table}[t]
    \centering
    \caption{Comparison of the number of factors of LZ78 and LZ78R on the introduced datasets. The first and the second column measures the number of factors of the respective compression scheme.
    The last column gives the percentage between the number of factors of LZ78R and LZ78, where more than 100\% means that the number of LZ78R factors is larger than of LZ78.
}
    \label{tabLZ78R}
\begin{tabular}{lrrr}
    dataset & LZ78 & LZ78R & \% \\
    \hline \Dataset{bib} & \num{21459} & \num{21573} & \num{100.53} \\
\Dataset{book1} & \num{131072} & \num{131090} & \num{100.01} \\
\Dataset{book2} & \num{102512} & \num{102422} & \num{99.91} \\
\Dataset{geo} & \num{26328} & \num{26314} & \num{99.95} \\
\Dataset{news} & \num{73434} & \num{73214} & \num{99.70} \\
\Dataset{obj1} & \num{6105} & \num{6035} & \num{98.85} \\
\Dataset{obj2} & \num{50905} & \num{50842} & \num{99.88} \\
\Dataset{paper1} & \num{12167} & \num{12197} & \num{100.25} \\
\Dataset{paper2} & \num{17337} & \num{17350} & \num{100.07} \\
\Dataset{paper3} & \num{10905} & \num{10918} & \num{100.12} \\
\Dataset{paper4} & \num{3649} & \num{3638} & \num{99.70} \\
\Dataset{paper5} & \num{3410} & \num{3412} & \num{100.06} \\
\Dataset{paper6} & \num{9149} & \num{9108} & \num{99.55} \\
\Dataset{pic} & \num{26646} & \num{26607} & \num{99.85} \\
\Dataset{progc} & \num{9459} & \num{9436} & \num{99.76} \\
\Dataset{progl} & \num{13624} & \num{13552} & \num{99.47} \\
\Dataset{progp} & \num{9812} & \num{9816} & \num{100.04} \\
\Dataset{trans} & \num{18200} & \num{18243} & \num{100.24} \\
\Dataset{alice29.txt} & \num{29091} & \num{29289} & \num{100.68} \\
\Dataset{asyoulik.txt} & \num{25591} & \num{25561} & \num{99.88} \\
\Dataset{cp.html} & \num{5685} & \num{5677} & \num{99.86} \\
\Dataset{fields.c} & \num{2785} & \num{2781} & \num{99.86} \\
\Dataset{grammar.lsp} & \num{1071} & \num{1065} & \num{99.44} \\
\Dataset{kennedy.xls} & \num{91430} & \num{91795} & \num{100.40} \\
\Dataset{lcet10.txt} & \num{72083} & \num{72286} & \num{100.28} \\
\Dataset{plrabn12.txt} & \num{84710} & \num{84676} & \num{99.96} \\
\Dataset{sum} & \num{8818} & \num{8913} & \num{101.08} \\
\Dataset{xargs.1} & \num{1344} & \num{1332} & \num{99.11} \\
 \end{tabular}
\end{table}

\subsection{Execution time and maximum memory usage}
We measured the execution time and maximum memory usage on the Pizza\&Chili corpus 
in \cref{fig:exectime,fig:maxmem}, respectively.
The execution time and maximum memory usage of all compression schemes appear to be linear in practice.
LZ78 and LZW from tudocomp seem to be the fastest compression algorithms as well as the ones with the lowest maximum memory usage.
LZ78 seems to be a bit faster than LZW in most cases, while the maximum memory usage has no clear winner between the both.
The LZDR, LZD+ and LZD algorithms have competing execution times, with LZD+ being slightly faster than LZDR, and LZDR being sometimes slower or faster than LZD\@.
The maximum memory usage for LZDR and LZD+ appears to be nearly identical, which is probably due to the algorithms being similar and using the naive LCE implementation for LZDR without extra LCE data structure.
LZD has a significantly lower maximum memory usage than both LZDR and LZD+ with more than a 500 MiB difference for the \textsc{english} dataset with an input size of 200 MiB.

These measurements evaluate our prototypes rather than the algorithms of \cref{thmLZDPlus,thmLZDR}: our LZD+ implementation is not tuned to the $O(z)$-word space bound, and our LZDR implementation computes LCE queries naively instead of using the LCE data structure assumed in \cref{thmLZDR}.
Our implementations of the flexible variants evaluate all lookahead alternatives explicitly and have no linear worst-case time bound.

\subsection{Repetition-Variant of LZ78}

It is also possible to define an LZ78 variant, which we call LZ78R, that uses the repetition rule like LZDR\@.
So when computing an LZ78 rule, we also track the longest repetition or truncation of a factor that can be used to represent the current factor.
Preliminary experiments listed in \cref{tabLZ78R} show however that LZ78R does not yield better results than the standard LZ78 variant.

\section{Conclusion}\label{sec:conclusion}
We introduced two new greedy compression schemes, LZD+ and LZDR\@.
LZD+ can be computed online in linear expected time with $O(z)$ words of working space, and LZDR offline in linear expected time with $O(n)$ words.
On the strings of Badkobeh et al., LZDR produces far fewer factors than LZD and LZD+ in our experiments; however, a family of binary strings shows that LZD, LZD+, and LZDR all have an approximation ratio of $\Omega(n^{1/3})$ with respect to the smallest grammar, while our bound in terms of the substring complexity limits it to $O((n/\log n)^{2/3})$.
Practical benchmarks show that both LZD+ and LZDR achieve a smaller factor count than LZD for small files at least, while the flexible parsing variants of LZDR usually seem to achieve the smallest factor count.
The compression times of LZDR, LZD+, and LZD are very similar in our benchmarks, though still slower than LZ78 and LZW.
The maximum memory usage for LZDR and LZD+ is higher than that of LZD, and LZ78 and LZW achieve the lowest maximum memory usage.
Further open research questions in regard to the newly introduced compression schemes are:
\begin{itemize}
  \item What would be a good way to store an LZDR or LZD+ parsing as a series of bytes? And how does the compressed file size of LZDR and LZD+ compare to other compression schemes like LZD?
  \item What could an online linear-time algorithm for LZDR look like, if one exists?
  \item What could a linear-time algorithm for a flexible parsing variant of LZDR look like, if one exists?
  \item How do flexible parsing variants for LZD+ compare in respect to factor count?
  \item 
Can the gap between $\Omega(n^{1/3})$ and $O((n/\log n)^{2/3})$ for the approximation ratio of LZD, LZD+, and LZDR be closed?

  \item 
Can greedy LZD+ or LZDR produce asymptotically more factors than LZD, despite their larger sets of candidates at each position?

  \item 
How efficiently can \StdFlex{}-LZDR and \MaxAlt{}-LZDR be decoded, given that they reference greedy factors not contained in their output?

\end{itemize}

\section*{Acknowledgements and Author Declarations}

We thank Tomasz Kociumaka for initial discussions about the compression sensitivity for LZD, and thoughts about making the string family $S_k$ binary.
Data is provided within the manuscript and available for download at \url{https://github.com/koeppl/lzdr-comp}.

The lower bound family $B_k$ was polished with AI assistance (openAI, Claude).

\section*{Funding Statements}
    This work was supported by JSPS KAKENHI Grant Number JP25K21150.

\bibliographystyle{plain}

\end{document}